\documentclass[twocolumn]{aastex62}

\usepackage{savesym}
\savesymbol{tablenum}
\usepackage{siunitx}
\restoresymbol{SIX}{tablenum}
\usepackage{amsmath}

\DeclareSIUnit\rg{r_g}

\newcommand{\rbp}{r$_{\rm BP}$}
\newcommand{\iin}{$i_{\rm in}$}
\newcommand{\iout}{$i_{\rm out}$}
\newcommand{\eq}[2]{#1 $=$ #2}
\newcommand{\apeq}[2]{#1 $\approx$ #2}

\graphicspath{{./}{figures/}}

\shorttitle{Bardeen-Petterson Polarization}
\shortauthors{Abarr \& Krawczynski}

\begin{document}

\title{The Polarization of X-rays from  Warped  Black Hole Accretion Disks}
\correspondingauthor{Quincy Abarr}
\email{qabarr@wustl.edu}

\author{Quincy Abarr}
\author{Henric Krawczynski}a
\affil{Washington University in St. Louis and McDonnell Center for the Space Sciences}

\begin{abstract}

It is commonly assumed that in black hole accretion disks the angular momenta of the disk and the black hole are aligned.
However, for a significant fraction of stellar mass black holes and supermassive black holes, the momenta may not be aligned.
In such systems, the interplay of disk viscosity and general relativistic frame dragging can cause the disk to warp or break into two (or more) distinct planes; this is called the Bardeen-Petterson effect. 
We have developed a general relativistic ray-tracing code to find the energy spectra and polarization of  warped  accretion disks, accounting for the emission from the disk and for photons reflecting one or multiple times off the  warped accretion disk segments. 
We find that polarization angle can be used to give a lower limit on the misalignment angle when a previous measurement of the jet, which is thought be aligned with the black hole angular momentum, can be spatially resolved.

\end{abstract}


\section{Introduction} \label{sec:intro}

The {\it Imaging X-ray Polarimetry Explorer (IXPE)} \citep{weisskopf2016} is scheduled for launch  in 2021 and will be the first dedicated satellite X-ray polarimeter since {\it OSO-8} was launched in 1978. 
In the meantime, a number of satellites and several balloon borne experiments have acquired first X-ray polarization information for a handful of the brightest X-ray sources. 
{\it IXPE} will be able to measure the soft X-ray (2-8 keV) polarization of X-ray binaries, Active Glactic Nuclei (AGNs), and supernova remnants including imaging
spectropolarimetric data for a handful of bright objects which
extend over more than 15'' in the sky. A complementary instrument, {\it XL-Calibur}, will fly on high altitude balloons for several missions from Sweden and Antarctica in 2021, 2022 and 2024. 
{\it XL-Calibur} will measure the polarization of 15-75 keV photons from accreting compact objects. The joint {\it IXPE} and {\it XL-Calibur}
observations will allow us to probe the astrophysics of the observed sources as well as general relativity and quantum electrodynamics (QED) in their respective strong field regimes.

Some of the scientific highlights of these missions will come from observations of accreting black holes (BHs) and neutron stars. 
X-ray polarimetry will build on the successes of the previous and current X-ray missions by providing the linear polarization fraction and angle as function of energy in addition to temporal, spectral and imaging information.  
The additional information will enable sensitive tests of the models developed and fine-tuned based on temporal, spectral, and imaging data alone.

Although the angular momenta of stellar mass black holes may be aligned with that of their binary orbits, the formation of these systems involve the explosion of a massive star and the strong kick from the supernova explosion may knock the momenta out of alignment. 
\citet{fragile2001} estimated that the median misalignment among post-supernova binaries is \SI{20}{\degree}, and that \SI{60}{\percent} of these binaries have a misalignment between \SIrange[range-phrase={ and }]{5}{45}{\degree}. 
For some systems the misalignment has been measured: it is estimated that the jet of GRO~J1655$-$40, likely aligned with the BH angular momentum, is inclined by \SI[separate-uncertainty=true]{85(2)}{\degree} \citep{hjellming1995} and the binary plane inclined by \SI[ separate-uncertainty=true ]{70.2+-1.9}{\degree} \citep{greene2001}. 
Galaxy merger are likely to produce supermassive BHs with misaligned accretion disks, depending on the details of the BH and galaxy mergers \citep{king2005}.

The impact of the accretion disk viscosity on a misaligned accretion disk was studied by \citet{BP1975}.
They posited that the Lense-Thirring (LT) effect \citep{lt1918} and the viscosity of the accretion disk material will align the disk with the BH angular momentum from the inner edge of the accretion disk outward.
More recent numerical work has found that in the thin disk regime where the viscosity $\alpha$ is greater than the aspect ratio $H/R$, the disk can warp or break into two or more distinct planes \citep{ogilvie1999, nixon2012a, nixon2012b}.
A disk broken into multiple sub-disks could accrete more efficiently due to LT precession cancelling the angular momentum of the individually precessing sections efficiently.  
This enhanced accretion rate could lead to flaring in the hard-intermediate state \citep{rm2006} and could also contribute to the rapid spin up of a BH, helping to explain the high spins of SMBHs \citep{risaliti2013}.

Previous simulations at moderate thickness of $H/R=0.08$ \citep[e.g.]{morales2018} did not find that the disk enters a BP configuration.
More recently, however, \citet{liska2019a} used the GRMHD code H-AMR \citep{liska2018} to simulate a very thin disk ($H/R=0.03$) with an initial tilt of \ang{10} surrounding a BH with spin \num{0.9375}.  
Their simulation included jets, which contributed to torque the inner disk into alignment \citep{mckinney2013} and found that their thin disk enters a persistent BP configuration with a transition radius \apeq{\rbp}{\SI{5}{\rg}} between the inner and outer disks.
At the end of the simulations, the inner disk had aligned with the BH angular momentum. The outer disk acquired an intermediate orientation
possibly due to the cancellation of misaligned angular momentum
owing to different precession angles of adjacent annuli  \citep{sorathia2013}.

The polarization of the accretion disk emission and the emission reflected off the disk should carry the imprint of the disk warp, if present. 
\citet{sk2009} showed that radiation being pulled down to the accretion disk by the gravity of the black hole and scattering off the disk has a stronger impact on the net polarization than on the energy spectrum. We expect that a disk warp will have an even stronger effect on the polarization, especially disk warps as close as \apeq{\rbp}{\SI{5}{\rg}} to the BHs as found by\citet{liska2019a}. 

In this work, we show the effect that a  warped  disk configuration similar to that of \citet{liska2019a} has on the polarization of the thermal emission from the accretion disk. 
We limit the discussion to a specific warp configuration viewed from different azimuthal viewing angles, and compare the observational signatures to those of the standard equatorial geometrically thin, optically thick disk. We improve on the work of \citet{cheng2016}, who studied the polarization of warped disks around a non-spinning and a spinning black hole with a misalignment of \SI{30}{\degree} and \rbp\, between \SIrange[range-phrase={ and }]{30}{1000}{r_g} in several ways.
We develop an approximate description of misaligned accretion disks in the Kerr metric allowing us to study the observational appearance of disks with warps in the very inner portion of the accretion flow 
where the background metric cannot be approximated by the 
Schwarzschild metric. As our code tracks photons forward in time, it allows us to model single and multiple reflections of photons off the
inner and outer portions of the accretion disk. In contrast \citet{cheng2016} included only single reflections of photons
from the inner disk reflecting off the outer disk.

Our code uses a simple geometry rather than the results from 
General Relativistic (Radiation) Magnetohydrodynamic (GR(R)MHD) simulations \citep[e.g.][]{liska2019a}. Such studies are important
for (i) interpreting the results of more involved GR(R)MHD simulations
and for (ii) giving us conceptual tools to interpret observational
results and discrepancies of observational and simulated results.  

The outline of the paper is as follows: Section \ref{sec:methods} gives a brief overview of the ray-tracing code we use. 
We briefly comment on the benefits of using the Cash-Karp method 
with adaptive step size to integrate the photon geodesics. 
We explain how we parameterize the misaligned disk, and how we implement photon reflections off the misaligned disk.
Section \ref{sec:results} compares the results for a warped disk to a standard equatorial disk, showing the impact of the warp on the flux and polarization energy spectra. 
In Section \ref{sec:discuss} we examine the results and study the impact of the azimuthal viewing angle, and the reflections off different disk sections on the flux and polarization spectra. 
We discuss the results and emphasize the opportunities to explore the dynamics of warped disks with the upcoming and future X-ray polarization measurements.

\section{Methods} \label{sec:methods}

\subsection{General Relativistic Ray-Tracing Code} \label{subsec:code}
Our ray tracing code \citep{krawcz2012,Hoor:16,beheshtipour2017,krawcz2019}) generates photon packages (or beams) in an accretion disk extending from the inner most stable circular orbit (ISCO) to $100r_g$. 
The code uses Boyer-Lindquist coordinates to describe the Kerr background spacetime, and integrates the geodesics of the photon packages emitted by the accretion disk and/or a point or spatially extended corona.
In this paper, we focus on photon packages thermally emitted by the accretion disk.
The photons are emitted in random directions and are weighted with the product of the 
radial brightness distribution for a geometrically thin, optically thick accretion disk of \citet{Page:74} and the limb brightening function of an indefinitely deep electron atmosphere \citet{chandrasekhar}.

We use the Cash-Karp method \citep{ck1990,schnittman2013} to integrate the geodesic equation:
\begin{displaymath}
    \frac{d^2x^\mu}{d\lambda^2}=-\Gamma^\mu_{\sigma\nu}\frac{dx^\sigma}{d\lambda}\frac{dx^\nu}{d\lambda},
\end{displaymath}
and to parallel transport the polarization vector $f^{\mu}$:
\begin{displaymath}
    \frac{dx^\mu}{d\lambda}=-\Gamma^\mu_{\sigma\nu}f^\sigma\frac{dx^\nu}{d\lambda}.
\end{displaymath}
The integration gives us the geodesics $x^{\mu}(t)$ and the photons' wavevector $k^{\mu}$ and polarization vector $f^{\mu}$ along the geodesics.

The Cash-Karp method is a 5th order Runge-Kutta numerical integration method which allows for an adaptive step size. 
This is especially important in general relativistic calculations since space is highly curved near the BH (where large step sizes introduce error into the photon's tracked values) and asymptotically flat far from the BH (where small step sizes take a proportionally enormous amount of CPU time to cross the distance between the disk and observer). 
The code keeps track of the integration error using $k^2$, which should be zero for a null geodesic, and $f^2$, which should be constant along the geodesic.

In our code, the step size varies by several orders of magnitude along a geodesic. Due to the tuneable accuracy, we can trade off accuracy versus computational cost. 
Everything else being equal, the Cash-Karp version runs about two times faster and has an order of magnitude improvement in tracking error
($\Delta=k^2$) over the 4th-order Runge-Kutta method we used previously \citep[e.g.]{krawcz2012}.

Photons are tracked until they get within \SI{0.02}{r_g} of the BH or until they reach \SI{10,000}{r_g}. When the latter occurs, the wave vector is transformed into the reference frame of a coordinate stationary observer for subsequent analysis.       

When a photon impinges on the accretion disk, it is scattered using the formalism of \citet{chandrasekhar} for scatterings of polarized photons reflecting off an infinitely thick electron scattering atmosphere. 
We assume a highly ionized accretion disk around a stellar mass BH
which reflects 100\% of the photons. The scattering is implemented in several steps: First, the wave vector and 
polarization four vector of the incoming beam are transformed from the global
Boyer-Lindquist coordinates into the Lorentz frame of the reflecting material. 
Given the polarization vector in this frame, the incoming Stokes parameters 
are calculated. A random scattered direction is drawn with equal probability per
solid angle, and Equation (164) and Table XXV from Chapter X of \citet{chandrasekhar}
are used to determine the Stokes parameters of the outgoing beam. 
The Stokes parameters are subsequently usedto calculate the polarization fraction and polarization angle of the outgoing beamand to calculate the statistical weight for the particular scattering direction.
In the final step, the polarization angle is converted into the local polarization vector, and the wave vector and polarization vector are transformed again into the global Boyer-Lindquist coordinates.
\subsection{Description of the Warped Accretion Disk} \label{subsec:bpdisk}
\begin{figure}[t]
    \centering
    \includegraphics[width=0.3\linewidth, angle=-90]{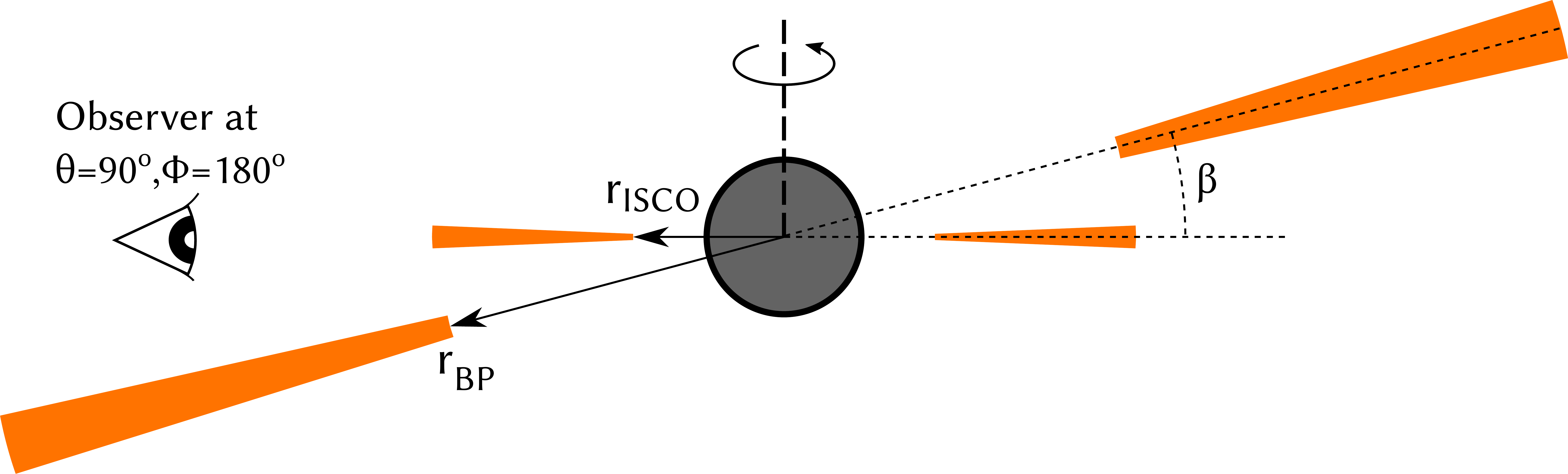}
    \caption{The warped disk configuration. The inner accretion disk is aligned with the BH spin axis. 
    At \rbp, there is a tilt of $\beta$ between the inner and outer disks. 
    Shown is an example observer at an inclination of \eq{\iin}{\ang{90}}, \eq{$\phi$}{\ang{180}}. 
    We use the spherical Boyer-Lindquist coordinates with \eq{$\theta$}{\ang{0}} pointing along the spin axis of the BH.
    For \eq{$\theta$}{\ang{90}}, \eq{$\phi$}{\ang{0}} points to the right, \eq{$\phi$}{\ang{180}} points to the left, and \eq{$\phi$}{\ang{270}} points to the reader.}
    \label{fig:bpdisk}
\end{figure}

We use the simplest version of a warped disk -- a broken disk residing on two planes.
For radial coordinates $r\le r_{\rm BP}$, we use a standard equatorial accretion disk
at a polar angle $\theta=\frac{\pi}{2}$ (see Fig.~\ref{fig:bpdisk}). 

In the Kerr metric, inclined test particle orbits precess and have 
cork-screw type shapes  \citep[e.g.][]{lt1918,wilkins72,liska2019a}. 
We assume that viscous stresses force the disk material onto orbits 
of constant $r$ for all $r>r_{\rm BP}$ inclined by an angle $\beta$ 
relative to the equatorial plane. 
The disk is defined by the solutions of the equation:
\begin{equation}
    \label{eq:disk}
    \cos(\theta)\cos(\beta)-\sin(\theta)\cos(\phi)\sin(\beta)=0.
\end{equation}
The disk thus extends from an inclination of $\theta=\ang{90}-\beta$ 
at $\phi=\ang{0}$ to $\theta=\ang{90}+\beta$ at $\phi-\ang{180}$ 
(Fig.~\ref{fig:bpdisk}).
The left-hand side of Eq. \ref{eq:disk} is positive above the disk and negative below, and so a photon
scatters off the inclined disk when this value changes sign. 

The Boyer-Lindquist coordinates of a disk segment are given by
\begin{equation}
    \label{eq:coord}
    x^{\mu}(t,r)\,=\,(t,r,\theta(t),\phi(t)).
\end{equation}
To find the angles $\theta$ and $\phi$, we first parameterize a circular orbit in the equatorial plane at radial coordinate $r$ around a BH with spin parameter $a\in\left[-1,+1\right]$.
This orbit has the Keplerian angular velocity 
\begin{equation}
    \Omega_{\rm K}\,=\,(a+r^{3/2})^{-1}.
    \label{eq:keplerian}
\end{equation}
Tilting the orbit by the angle $\beta$ gives
\begin{eqnarray}
    \label{eq:coord_angles}
    \theta & =  \arccos\left(\sin\beta\cos\left(\Omega_{\rm K} t \right) \right)\\
    \phi & =  \arctan\left( \sec\beta \tan\left(\Omega_{\rm K} t\right) \right).
\end{eqnarray}

The reflection off the inclined disk requires the transformation of the beam's wave vector and polarization vector from the global Boyer-Lindquist coordinates into the rest frame of the reflecting material.
For this purpose, we define a tetrad (system of orthonormal basis vectors) associated with the reflecting material, denoting the basis vectors  associated with the Boyer-Lindquist coordinates as $\partial_{\mu}$ and the new tetrad basis vectors as $\mathbf{e}_{\hat{a}}$ with $a\,=$ 0, 1, 2, 3.  
The first basis vector $\mathbf{e}_{\hat{0}}$ is simply the four velocity of the disk material $\mathbf{u}\,=\,dx^{\mu}/d\tau$.
The four velocity is proportional to $dx^{\mu}/dt$ but is normalized to -1. 
We choose $\mathbf{e}_{\hat{1}}$ and $\mathbf{e}_{\hat{3}}$ to be tangent to the inclined disk: $\mathbf{e}_{\hat{1}} \propto \partial_r$, and $\mathbf{e}_{\hat{3}}$ is the component of vector tangent to the particle orbit perpendicular to $\mathbf{e}_{\hat{0}}$ and $\mathbf{e}_{\hat{1}}$.
The last basis vector $\mathbf{e}_{\hat{2}}$ is the component of the gradient across the disk perpendicular to the first three basis vectors.
For $\mathbf{e}_{\hat{2}}$ and $\mathbf{e}_{\hat{3}}$ we get the perpendicular components to the other basis vectors by using Gram-Schmidt orthonomalization to subtract out the parallel components with the help of the metric.

In terms of the Boyer-Lindquist basis vectors $\partial_{\mu}$, 
the new basis vectors are given:
\begin{widetext}
\begin{eqnarray} \label{eq:bpbasis}
    \mathbf{e}_{\hat{0}} =& \left( \partial_t + \Omega_{\rm K} \sin\beta \sin\phi \partial_{\theta} + \Omega_{\rm K} \cos\beta \csc^2\theta \partial_{\phi} \right) / \sqrt{\Sigma}\\ \nonumber
    \mathbf{e}_{\hat{1}} =& \partial_{r} / \sqrt{g_{rr}} \\ \nonumber
    \mathbf{e}_{\hat{2}} =& \frac{ \sqrt{g_{\theta\theta}} \sin\beta \sin\phi }{ \sqrt{\Pi \left(g_{t\phi}^2 - g_{tt} g_{\phi\phi} \right)}} \left( g_{t\phi} \partial_t - g_{tt} g_{\phi\phi} \partial_\phi \right) - \sqrt{ \frac{ g_{t\phi}^2 - g_{tt} g_{\phi\phi} }{ \Pi g_{\theta\theta} }} \cos\beta \csc^2\theta \partial_\theta \\
    \mathbf{e}_{\hat{3}} =&\left[ \left( -g_{t\phi} \cos\beta \csc^2\theta - g_{\phi\phi} \Omega_{\rm K} \cos^2\beta \csc^4\theta - g_{\theta\theta} \Omega_{\rm K} \sin^2\beta \sin^2\phi \right) \partial_t \right. \nonumber \\
                          &\qquad \left. + \left( g_{tt} + g_{t\phi} \Omega_{\rm K} \cos\beta \csc^2\theta \right) \left( \sin\beta \sin\phi \partial_{\theta} + \cos\beta \csc^2\theta \partial_{\phi} \right) \right] \nonumber \\
                         &\qquad\div \sqrt{ \Sigma \cos^2\beta \csc^4\theta \left( g_{t\phi}^2 - g_{tt} g_{\phi\phi} \right) + g_{tt} g_{\theta\theta} \sin^2\beta \sin^2\phi} \nonumber
\end{eqnarray}
where
\begin{eqnarray*}
    \Sigma =& -g_{tt} - \Omega_{\rm K} \left( 2g_{t\phi} \cos\beta \csc^2\theta + g_{\phi\phi} \Omega_{\rm K} \cos^2\beta \csc^4\theta + g_{\theta\theta} \Omega_{\rm K} \sin^2\beta \sin^2\phi \right) \\
    \Pi =& \left( g_{t\phi} ^2 - g_{tt} g_{\phi\phi} \right) \cos^2\beta \csc^4\theta - g_{tt} g_{\theta\theta} \sin^2\beta \sin^2\phi.
\end{eqnarray*}
\end{widetext}
 
The transformations can be effected by dotting a four vector in Boyer-Lindquist coordinates with the four basis vectors (giving, up to a sign for the zero-component, the coordinates in the accretion disk frame), and by multiplying the accretion disk coordinates with the respective basis vectors in Boyer-Lindquist coordinates.

Eq. \ref{eq:bpbasis} is only valid in the upper hemisphere of the inclined disk, so we restrict photon emission to $k^\theta<0$. It is possible that a photon emitted from the inner disk near $\phi=0^\circ$ scatters off the bottom of the outer disk, or a photon emitted from the outer disk near $\phi=180^\circ$ scatters off the bottom of the inner disk, in which case the reflected photon is emitted back into the lower hemisphere. In this case, we take advantage of the symmetry of the disk to only scatter photons into the upper hemisphere: we "mirror" the reflected photon with 
\begin{eqnarray*}
    \theta \to \pi-\theta \\
    \phi \to \phi+\pi \\
    k^\theta \to -k^\theta \\
    f^\theta \to -f^\theta
    \label{eq:mirror}
\end{eqnarray*}
This changes our tracked photon into the equivalent photon which was initially emitted into the \textit{lower} hemisphere and scattered into the \textit{upper} hemisphere. 
During analysis, we perform this same transformation to any photons arriving in the lower hemisphere to account any of these equivalent photons ending up in the upper hemisphere.

\section{Results} \label{sec:results}

In this study, we focus on a single warped disk configuration around a stellar mass BH.
We choose a BH with a spin of 0.9, \rbp\, of \SI{8}{\rg}, and a misalignment of \ang{15}. 
The BH has a mass of \SI{10}{M_{\odot}} and an accretion rate of \SI{8.98e18}{\gram/\second}, or \SI{0.5}{\dot{M}_{\rm E}}.

We generate 3.5$\times 10^8$ photon packages between $r_{\rm ISCO}$\SI{=2.32}{r_g} and \SI{100}{r_g} for the  warped disk, and compare the results to those for $10^7$ photons generated for a standard equatorial disk. 
The former case requires more photons due to the lack of azimuthal symmetry. 
For each observer, we collect all photons within \ang{4} of the location of the observer.

We will discuss in the following the observers at a constant inclination
of \eq{\iin}{\ang{75}} measured from the angular momentum vector of the black hole and the inner disk. 
\newcommand{\ob}[1]{\textbf{Ob#1}}
In Table \ref{tab:observers} we list the eight observers analyzed, each with a different azimuthal angles $\phi$ starting at \eq{$\phi$}{\ang{90}} (\ob{90}). Given $i$, $\phi$, and $\beta$ the inclination of the outer disk is given by:
\begin{equation}
    i_{\rm out}=\arccos\left( \cos{i_{\rm in}} \cos{\beta}-\sin{i_{\rm in}} \cos\phi \sin\beta \right)
    \label{eq:iprime}
\end{equation}
with \iin being the inclination of the inner disk.
Thus, for observer {\bf Ob90}, the outer disk is inclined at \eq{\iout}{\ang{75.52}}, almost the same as \iin.

\begin{figure*}
    \gridline{\fig{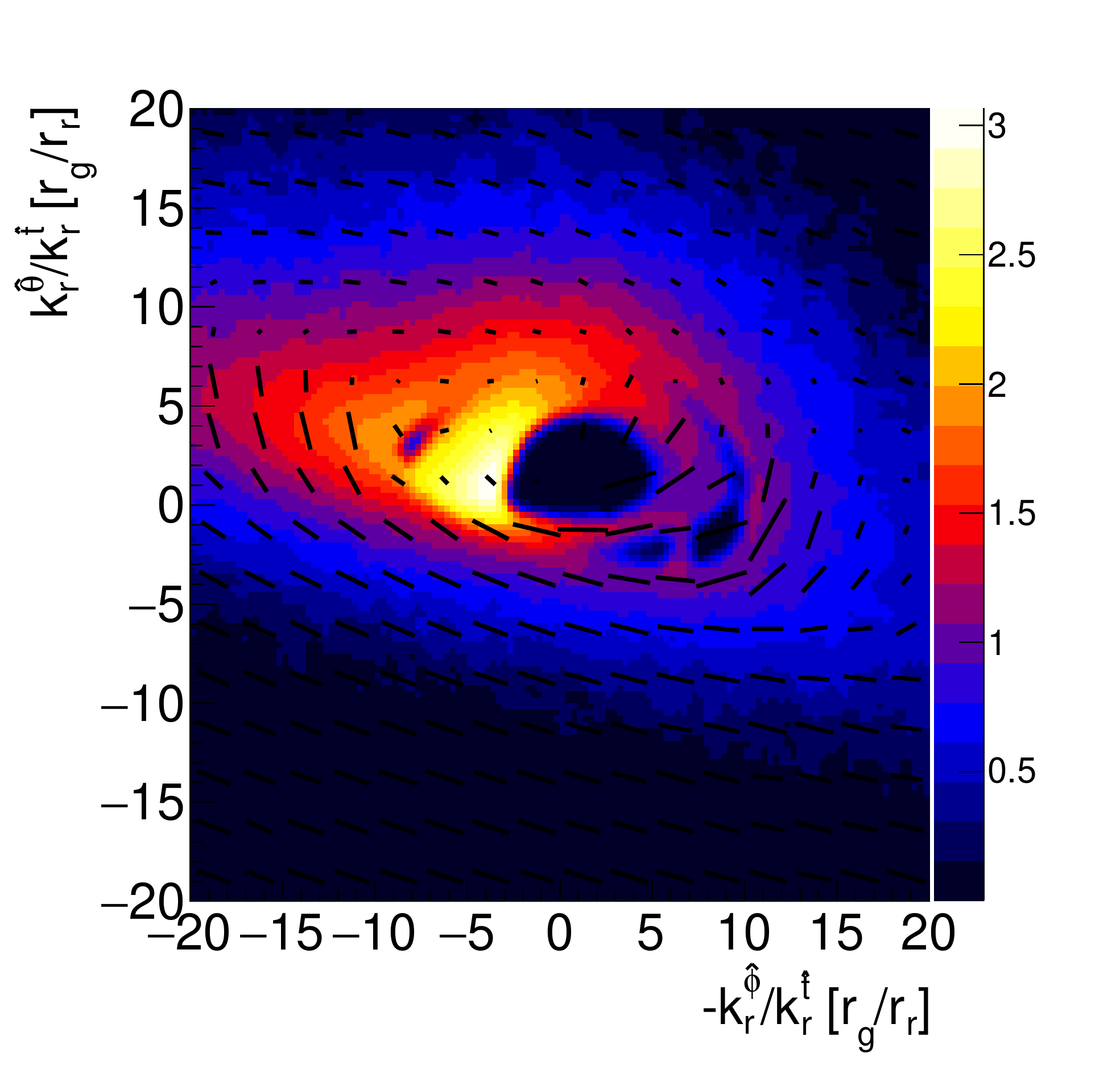}{0.33\linewidth}{}
              \fig{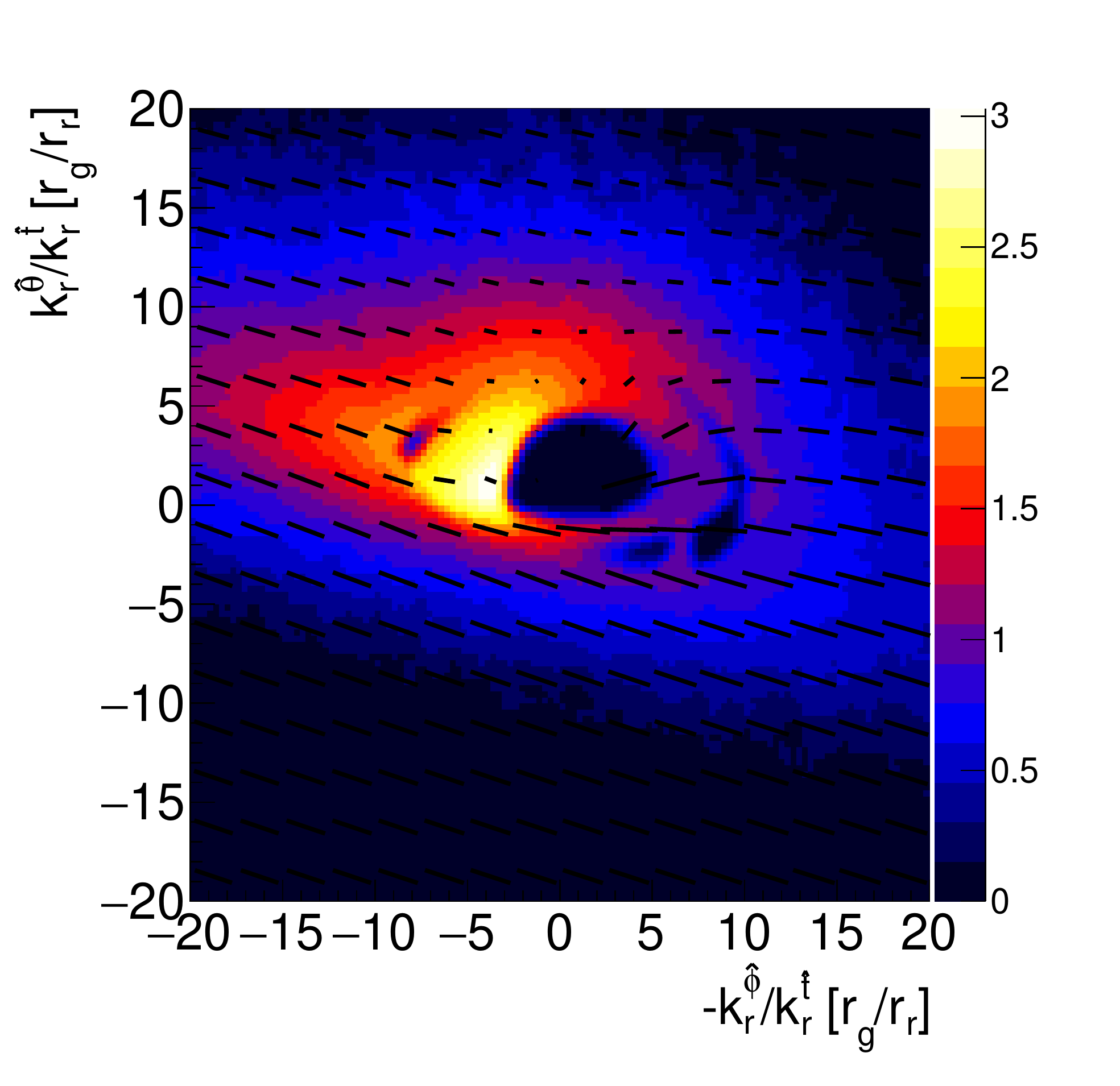}{0.33\linewidth}{}
              \fig{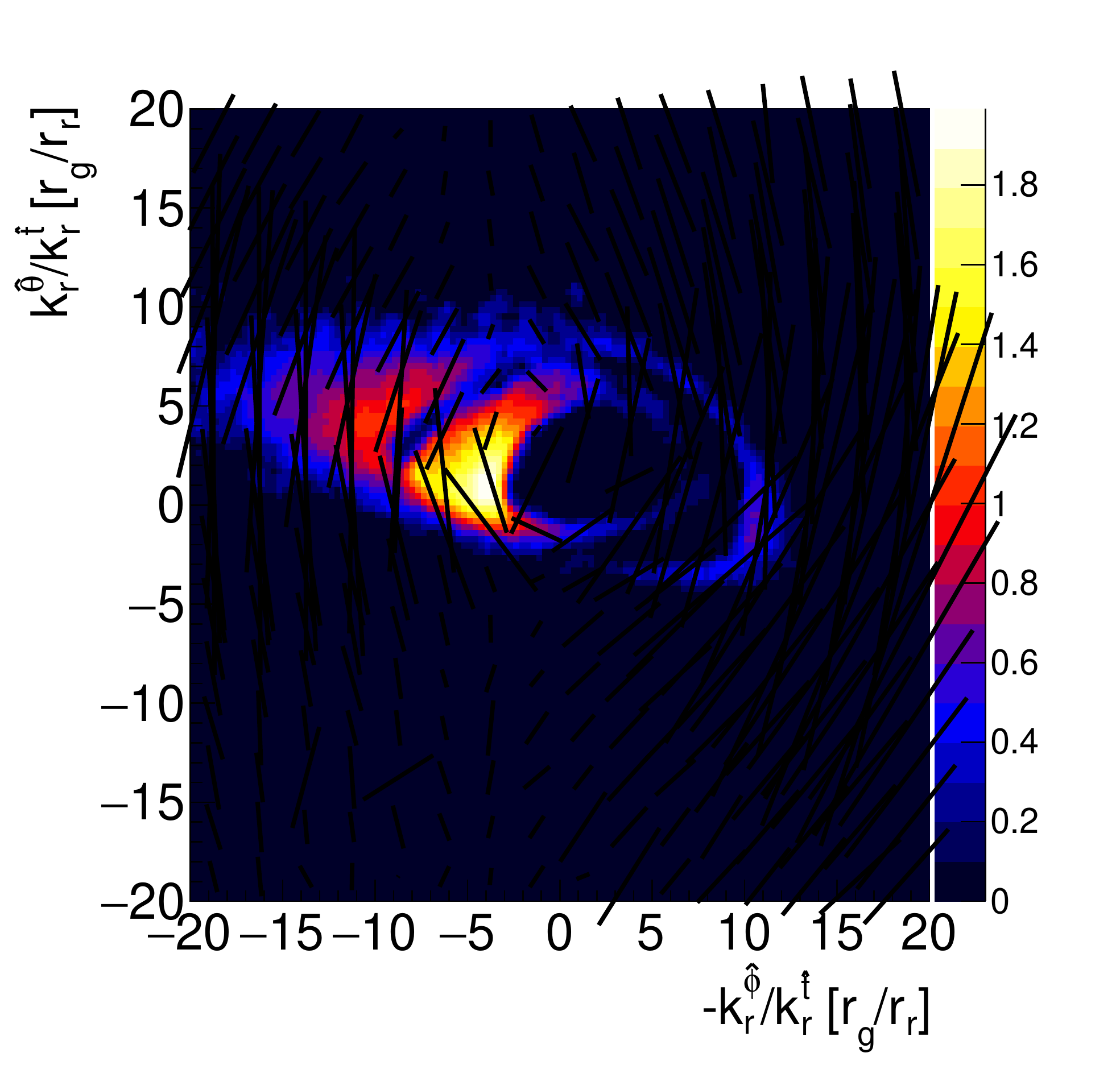}{0.33\linewidth}{}}
              \caption{The three emission types we examine: Total emission (left), direct emission (middle), reflected emission (right).
              The color bar gives the surface brightness in logarithmic units. 
              Over the images we have plotted the polarization, where the length (orientation) of the black bars gives the polarization fraction (angle). }
    \label{fig:images90}
\end{figure*}

We can distinguish between three classes of events: total emission, direct emission, and reflected emission off the whole disk.
Figure \ref{fig:images90} shows BH images for the three event classes, with their polarization plotted on top.
We see that the direct emission (middle panel) is polarized along the plane of disk from which it emits; this is more obvious for the outer disk, where the emission is less affected by light bending
The front of the inner disk is more horizontally polarized than the outer disk, and its back is more weakly polarized since the strong light bending causes this part of the disk to be viewed at a lower inclination angle.
The reflected emission (right panel) shows high degrees of polarization, and tends to come from the region close to the black hole.
On the left side of the black hole, where emission is beamed towards the observer, we see that photons can reflect at radii of up to \SIrange[range-phrase={ or }]{15}{20}{\rg}; for a fully aligned disk, we expect almost all scattering of returning radiation to occur within \SI{10}{\rg}.

\begin{figure*}
    \gridline{\fig{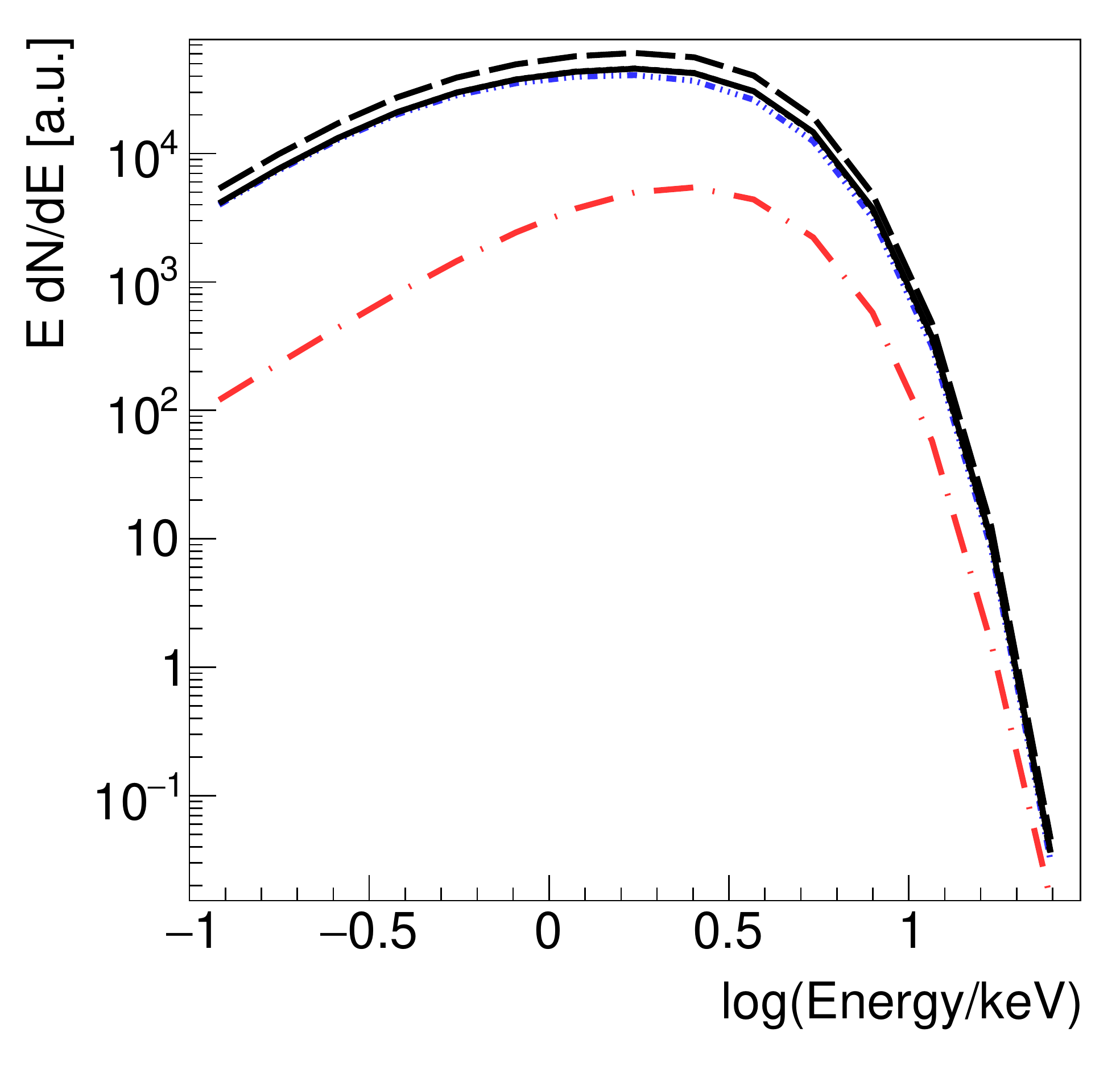}{0.3\linewidth}{}
              \fig{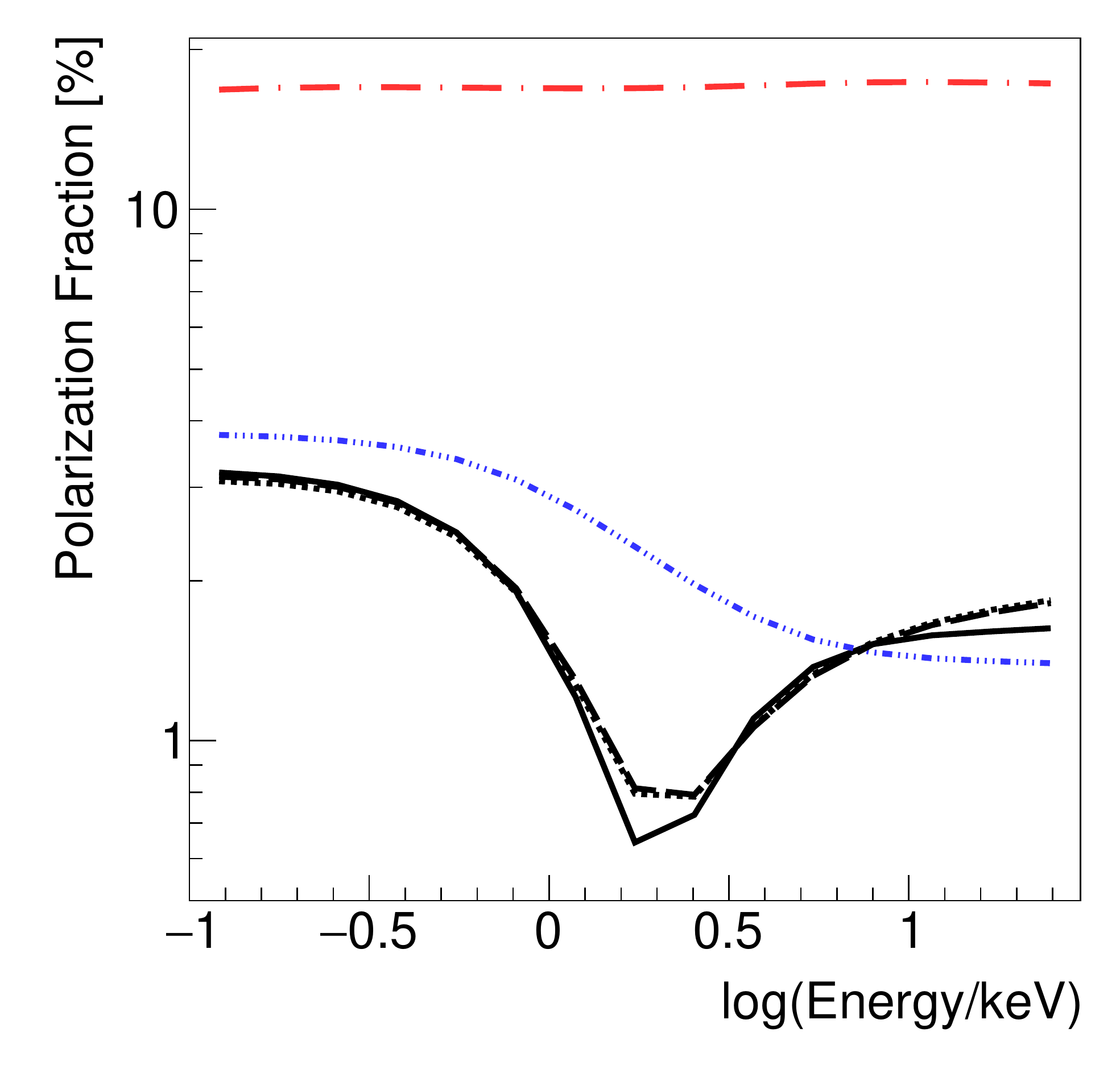}{0.3\linewidth}{}
              \fig{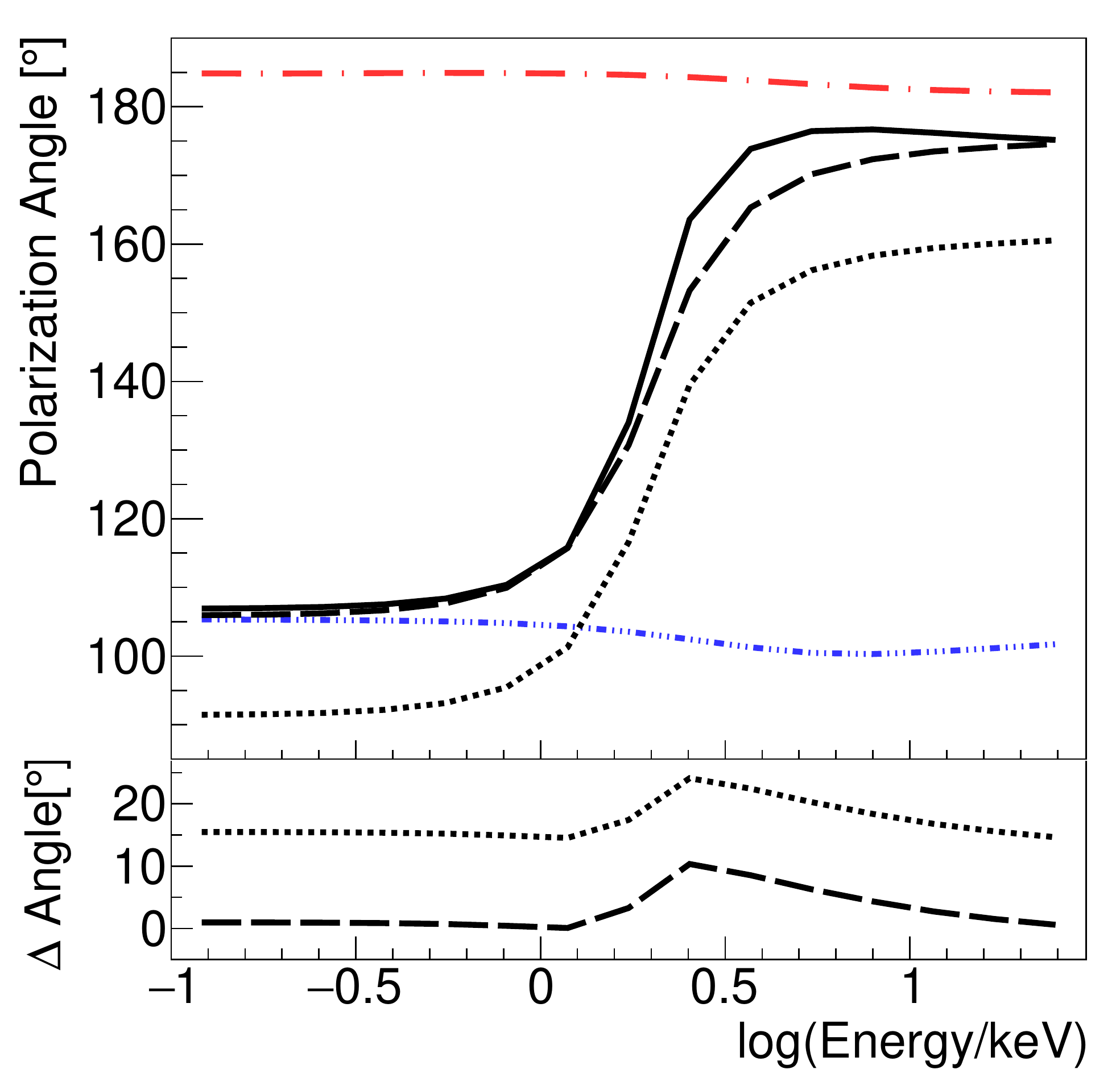}{0.3\linewidth}{}}
    \caption{Energy spectrum (left), polarization fraction (middle), and polarization angle (right) for the warped disk simulations ($r_{BP}\,=$ $8\,r_{\rm g}$, $\beta\,=$ $15^{\circ}$) and selected results from the equatorial disk for an observer at  $\theta\,=$ 75$^{\circ}$ and $\phi=90^{\circ}$. Included are the total emission (thick black), direct emission (dash-dot-dotted blue), and reflected emission (dash-dotted red). These are compared to the total emission of the completely aligned (equatorial) disk (dotted black) and the completely misaligned disk (dashed black).}
    \label{fig:phi90}
\end{figure*}

In Figure \ref{fig:phi90}, we show the energy spectrum, polarization fraction, and polarization angle for the observer \ob{90} compared to the total emission from two different flat disks: the completely aligned disk lies in the equatorial plane of the black hole, and the completely misaligned disk lies in the plane of the binary orbit (the same plane as the warped outer disk).
It is important to note that for both the completely aligned disk and the completely misaligned disk the angular momenta of the BH and disk are aligned; these are only "misaligned" in the sense that they represent unwarped disks in the same plane as the inner and outer disks, respectively.
We see that the flux and polarization fraction of the warped disk are very similar to those of the completely aligned disk.
The polarization angle of the warped disk configuration, however, is shifted by $\sim15^\circ$ and roughly matches the angle of the completely misaligned disk. 
It matches particularly well at low energies, where direct emission from the outer disk dominates, and seems to settle back down at high energies where the reflected emission dominates.

\begin{figure}[b]
    \centering
    \includegraphics[width=0.49\linewidth]{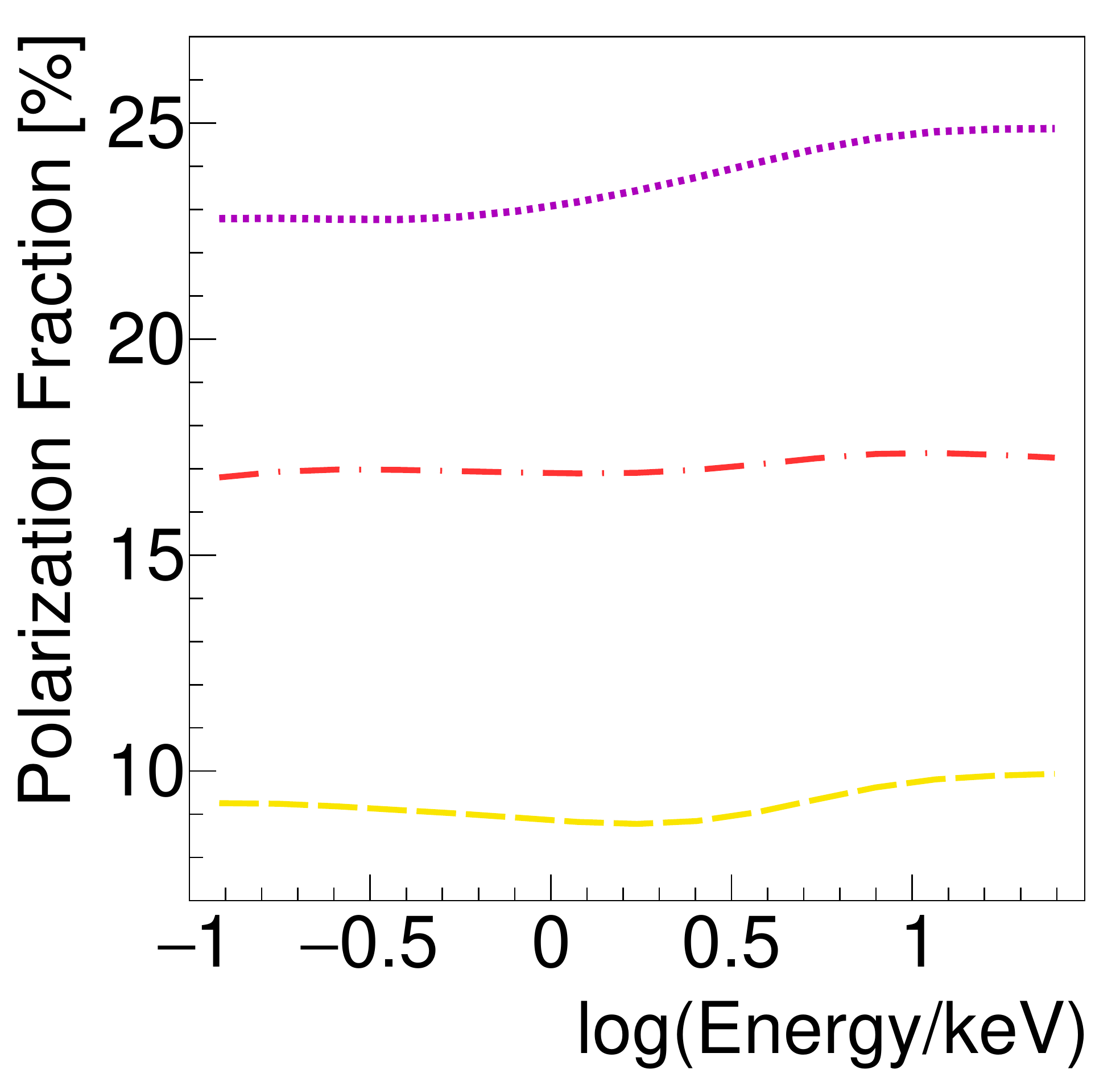}
    \includegraphics[width=0.49\linewidth]{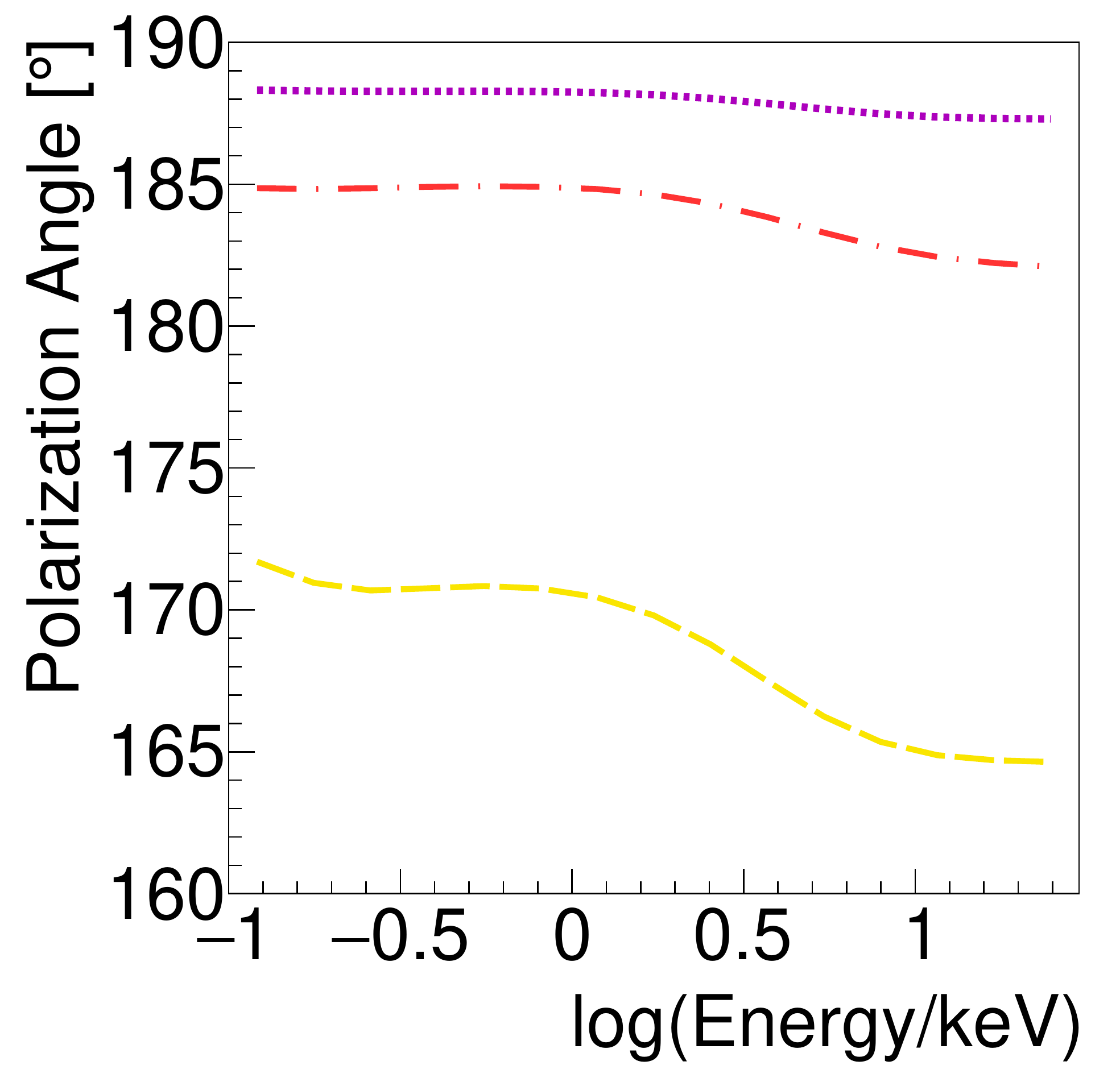}
    \caption{Polarization of the reflected emission in Fig. \ref{fig:phi90} split up into the portion reflected only off the inner disk (dashed yellow) and only off the outer disk (dotted purple).}
    \label{fig:pol_refl}
\end{figure}

In Figure \ref{fig:pol_refl}, we split the reflected emission into two parts: emission reflected off the outer disk and emission reflected off the inner disk. 
These two are exclusive, as there is a third (small) component of the reflected emission that reflects off both disks at least once. 
The overall reflected polarization fraction and angle clearly falls between the two components, with the outer disk component being more highly polarized than the inner disk.

\begin{table}[]
    \centering
    \begin{tabular}{|c|c|c|c|}
        \hline
        Observer & $\phi$-viewing angle & \iout & Figure \\ \hline \hline
        \ob{90} & \ang{90} & \ang{75.52} &  \ref{fig:images90}, \ref{fig:phi90}  \\ \hline
        \ob{135} & \ang{135} & \ang{64.74} &  \ref{fig:phi135}  \\ \hline
        \ob{180} & \ang{180} & \ang{60} &  \ref{fig:phi180}  \\ \hline
        \ob{225} & \ang{225} & \ang{64.74} &  Not included  \\ \hline
        \ob{270} & \ang{270} & \ang{75.52} &  Not included  \\ \hline
        \ob{315} & \ang{315} & \ang{85.8} &  \ref{fig:phi315}  \\ \hline
        \ob{0} & \ang{0} & \ang{90} &  \ref{fig:phi0}  \\ \hline
        \ob{45} & \ang{45} & \ang{85.8} &  Not included \\ \hline
    \end{tabular}
    \caption{Summary of the eight observers we use in this work}
    \label{tab:observers}
\end{table}

In the following, we discuss how the signatures change with the azimuthal viewing angle $\phi$.
We analyzed the eight observers listed in Table \ref{tab:observers}, but we will only discuss observers which are representative of the changes we see over changing $\phi$.  

\begin{figure*}[t]
    \gridline{\fig{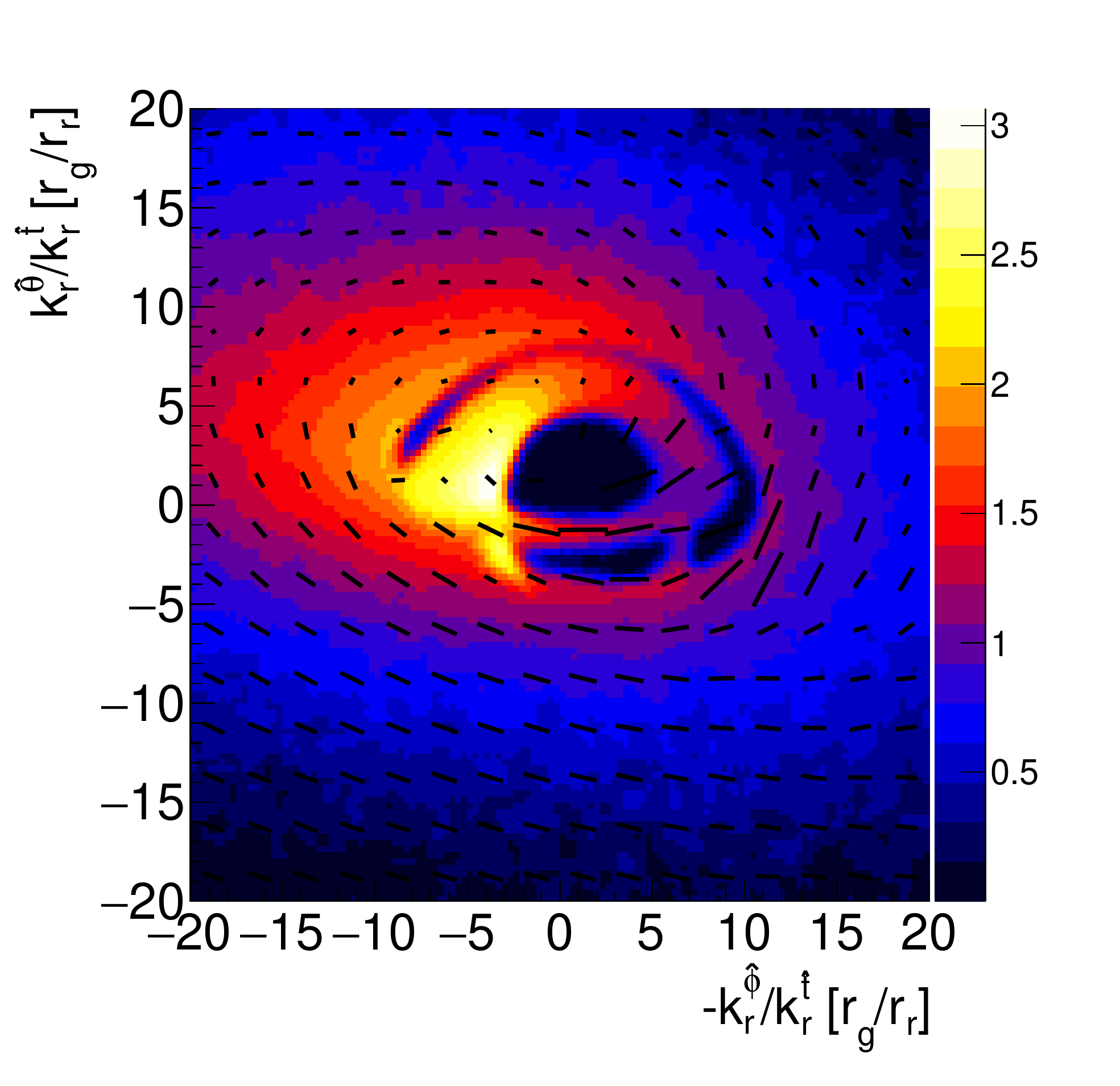}{0.25\linewidth}{}
              \fig{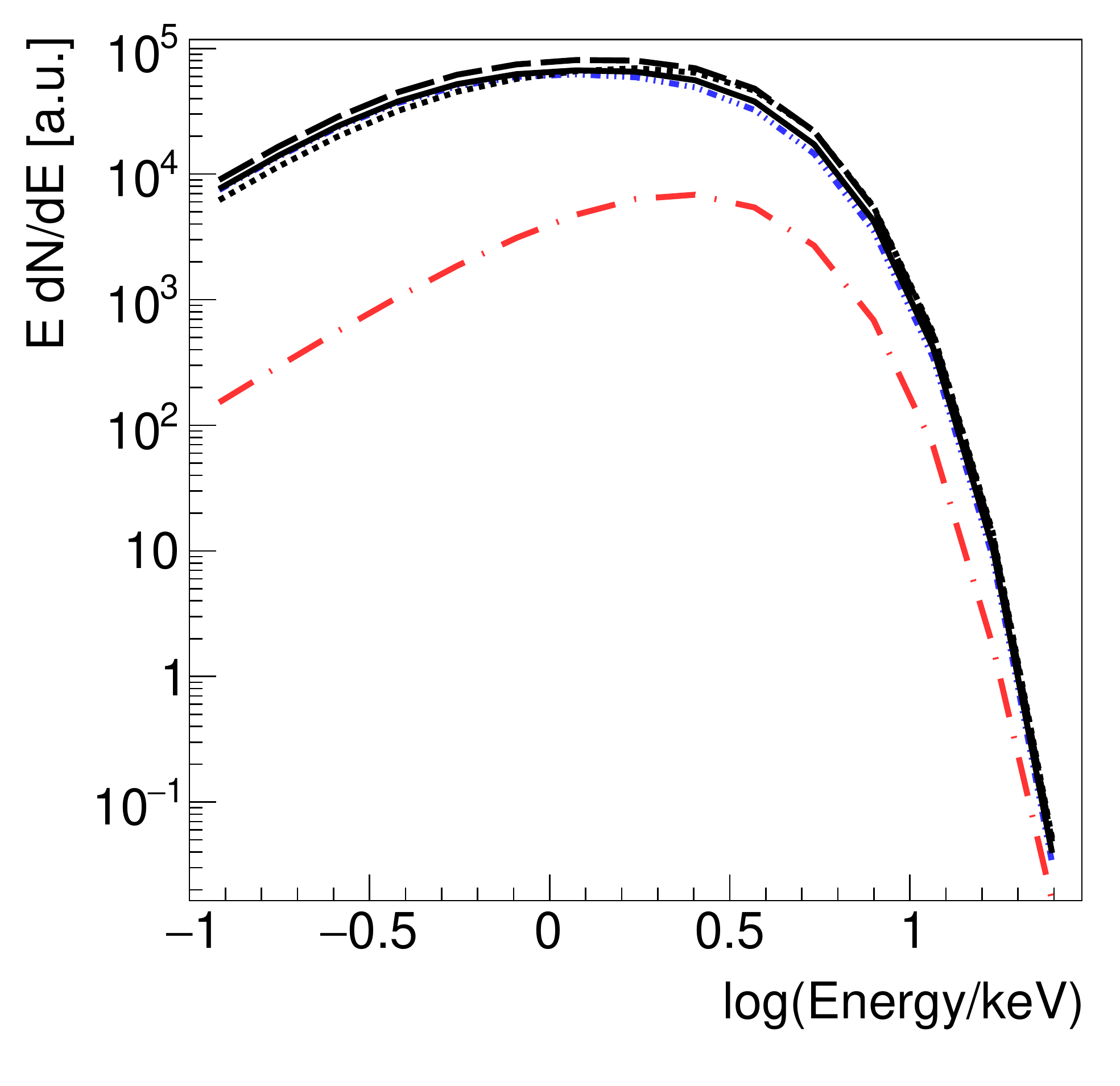}{0.25\linewidth}{}
              \fig{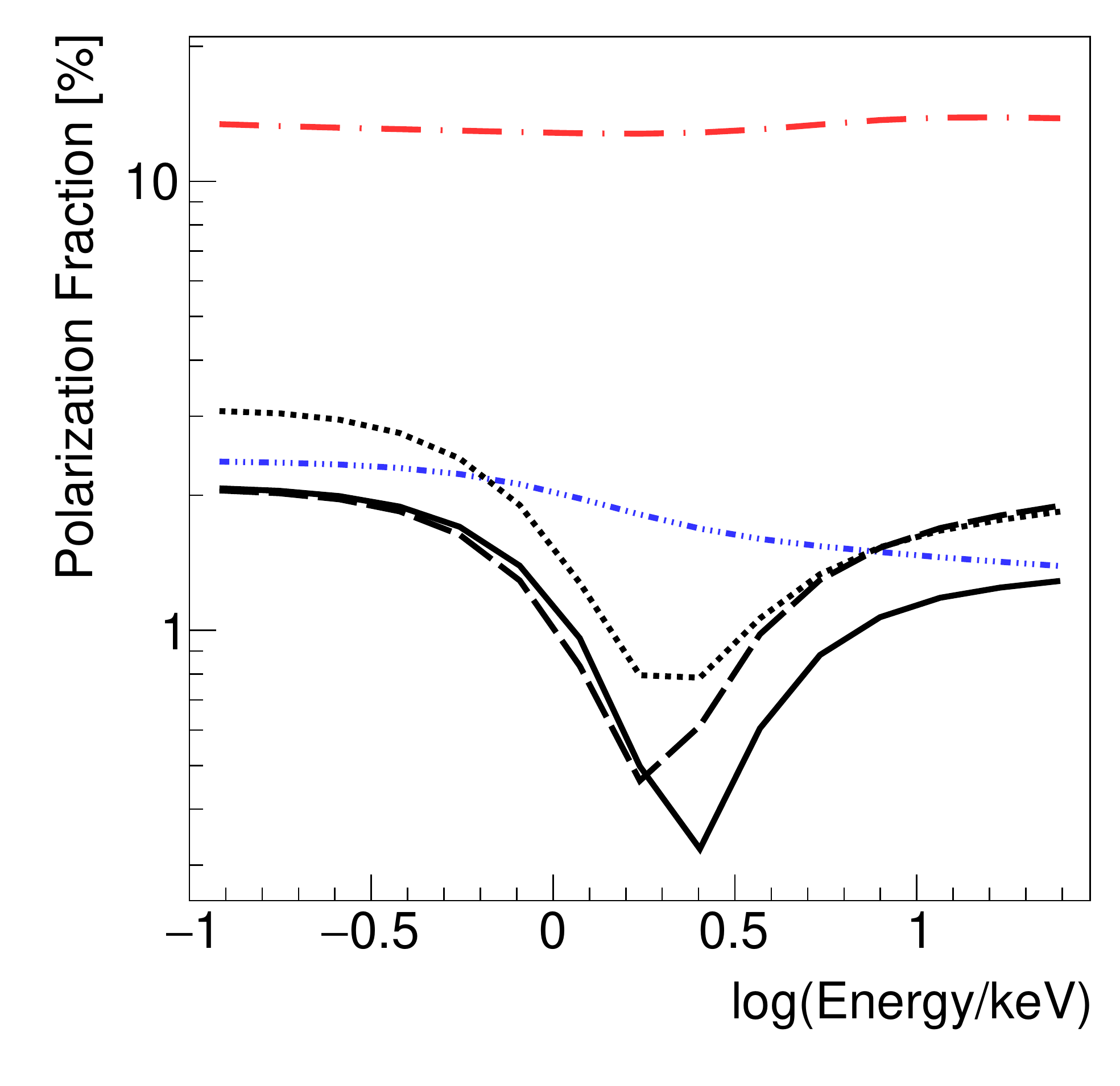}{0.25\linewidth}{}
              \fig{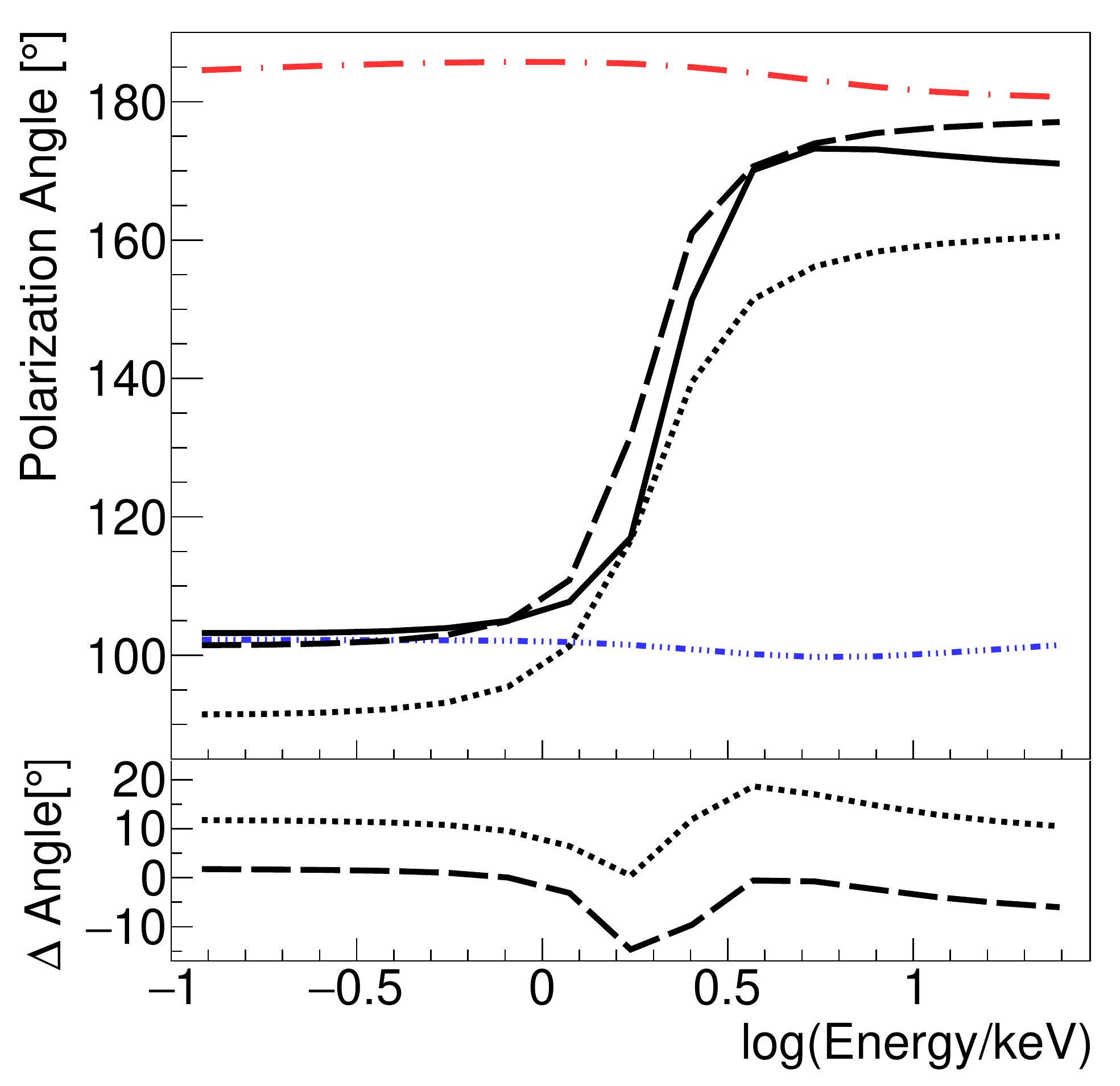}{0.25\linewidth}{}}
    \caption{Image of the BH (far left), energy spectrum (middle left), polarization fraction (middle right), and polarization angle (far right) seen by \ob{135}. Included are the total emission, direct emission, and reflected emission of the warped disk, as well as the total emission of the completely aligned and completely misaligned disks. This uses the same style conventions as in Figure \ref{fig:phi90}. For clarity, the bottom panel of the polarization angle plot shows the difference between the warped disk and the flat disks.}
    \label{fig:phi135}
\end{figure*}

\ob{135} (Fig.\ \ref{fig:phi135}) sees a slightly softer energy spectrum since more of the outer disk is visible and thus more direct emission is reaching the observer.
The polarization fraction is slightly lower than in the completely aligned disk.
At low energies, though, it matches the completely misaligned disk well.
The polarization angle roughly matches the completely misaligned disk, which is offset from the completely aligned disk by \ang{\sim12}; this value is the angle of misalignment visible to the observer.

\begin{figure*}[t]
    \gridline{\fig{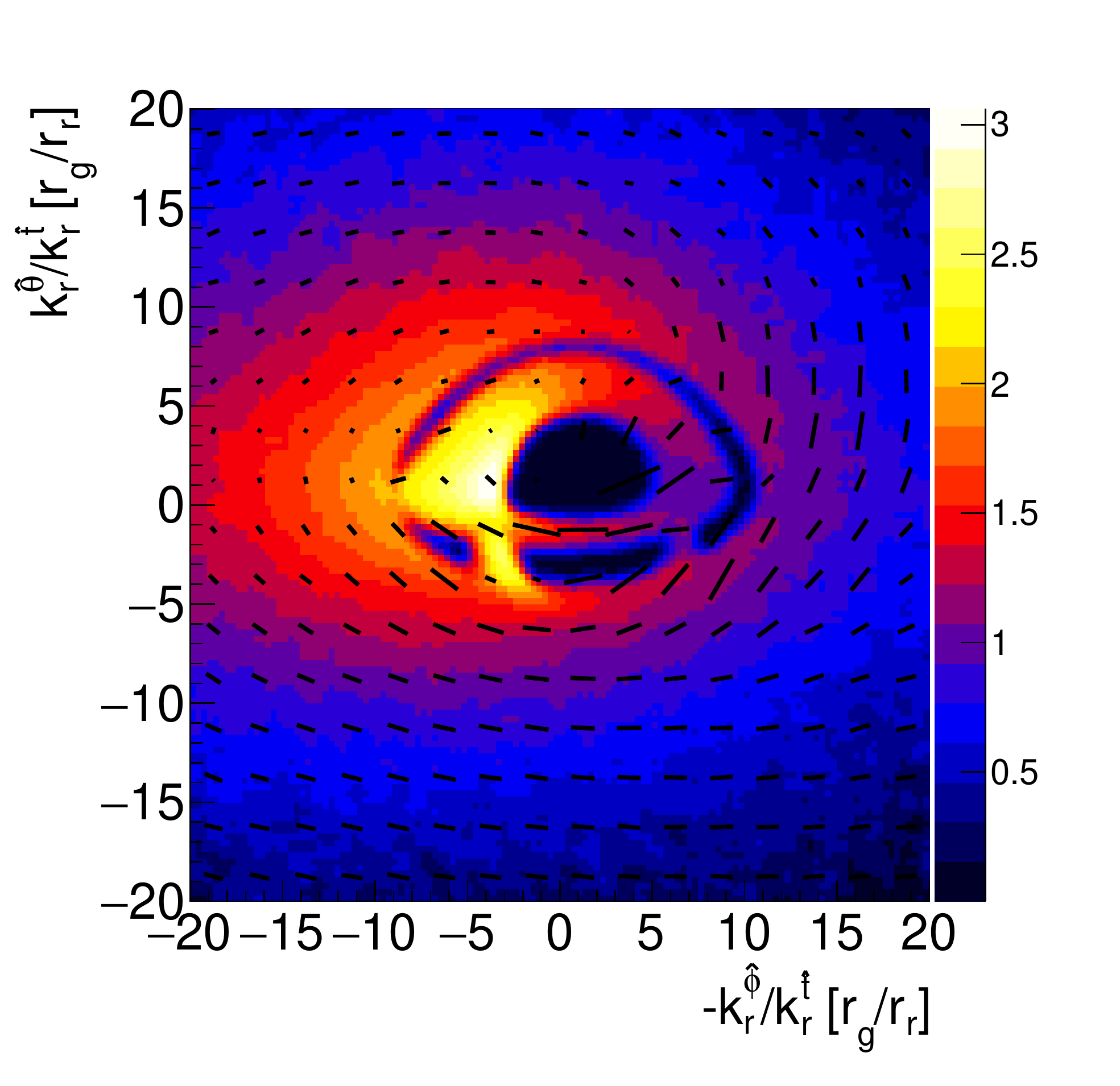}{0.25\linewidth}{}
              \fig{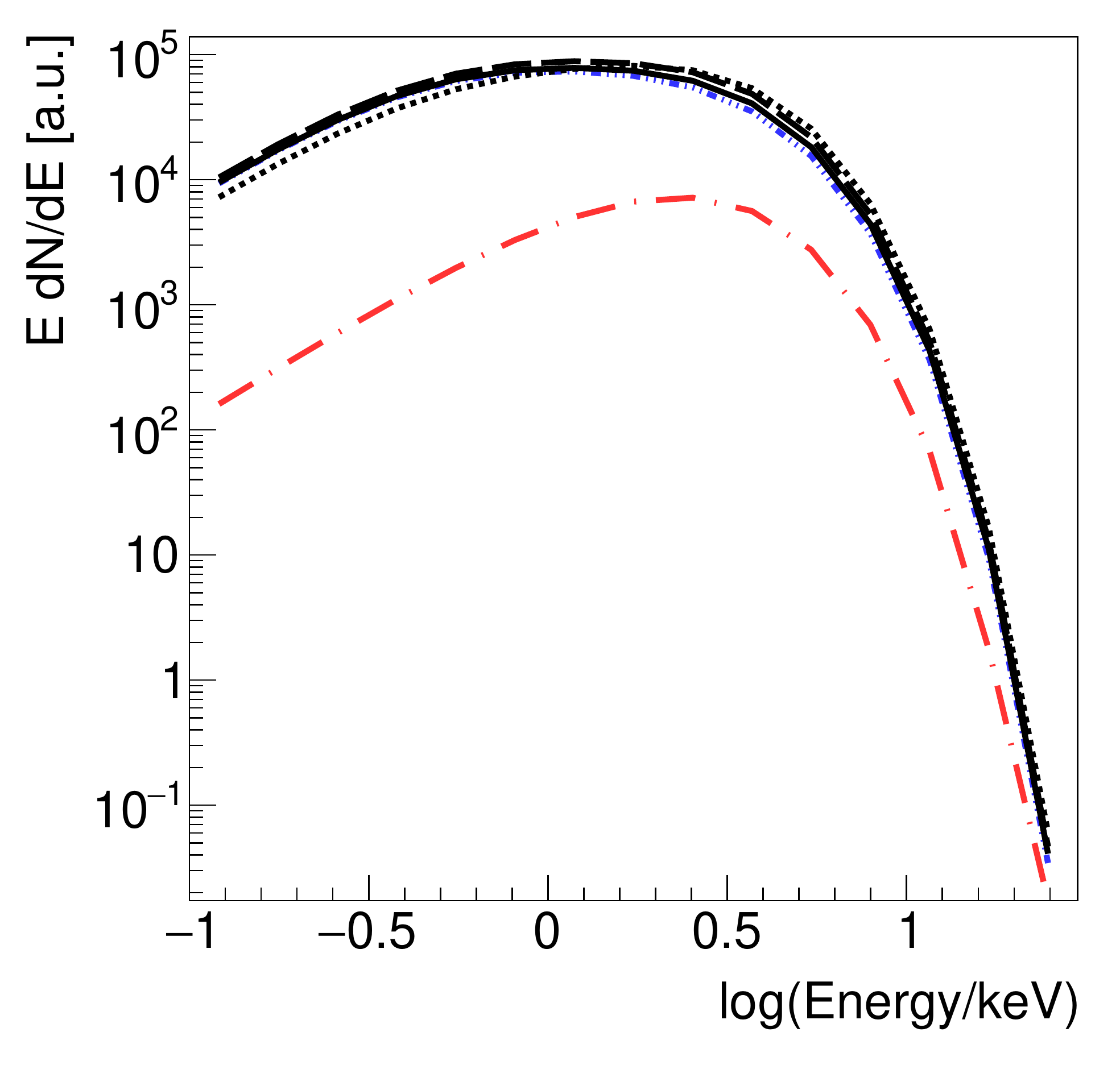}{0.25\linewidth}{}
              \fig{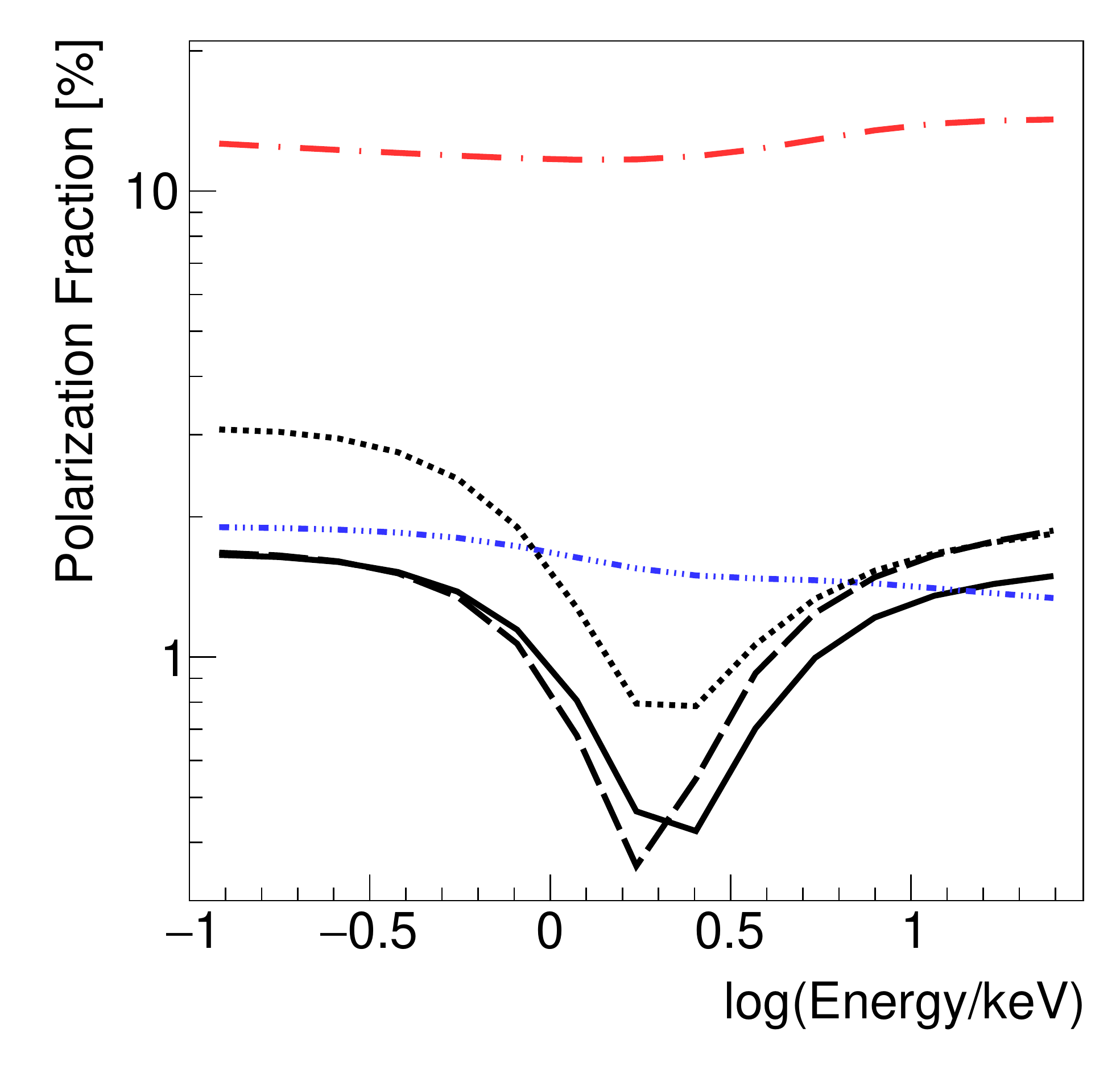}{0.25\linewidth}{}
              \fig{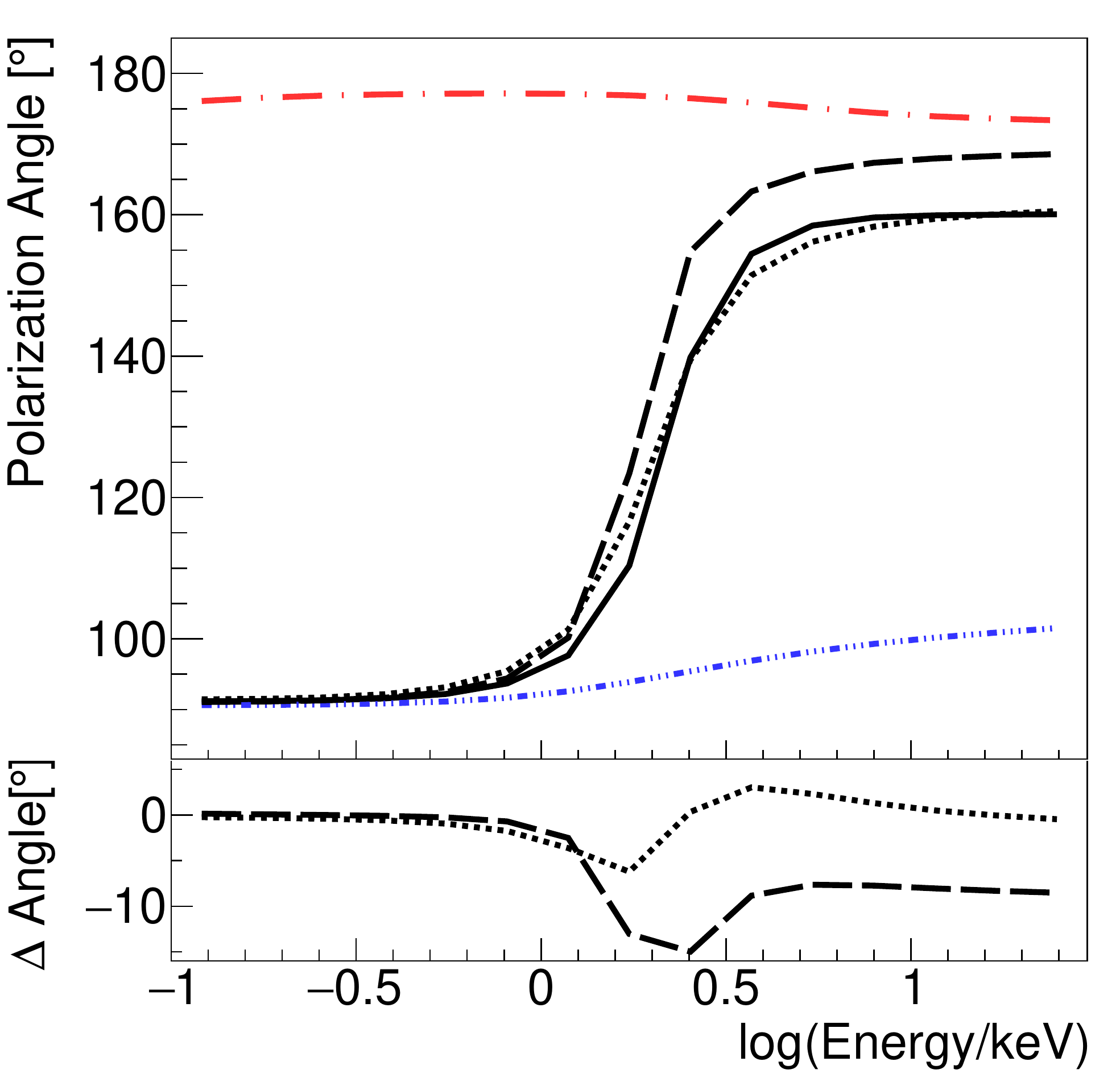}{0.25\linewidth}{}}
    \caption{As in Figure \ref{fig:phi135}, but for \ob{180}.}
    \label{fig:phi180}
\end{figure*}

Qualitatively, the flux and polarization fraction seen by \ob{180} (Fig.\ \ref{fig:phi180}) match that seen by \ob{135} (Fig.\ \ref{fig:phi135}). 
The polarization fraction, though, matches the completely aligned disk rather than the completely misaligned disk.
\citet{cheng2016} reported that the warped disk polarization angle matched that of their completely aligned disk.
Their system orientation was limited to one similar to our \ob{180} (Fig.\ \ref{fig:phi180}), though; the previously examined observers make it clear that in general the polarization angle doesn't match the completely aligned disk.

\begin{figure*}[t]
    \gridline{\fig{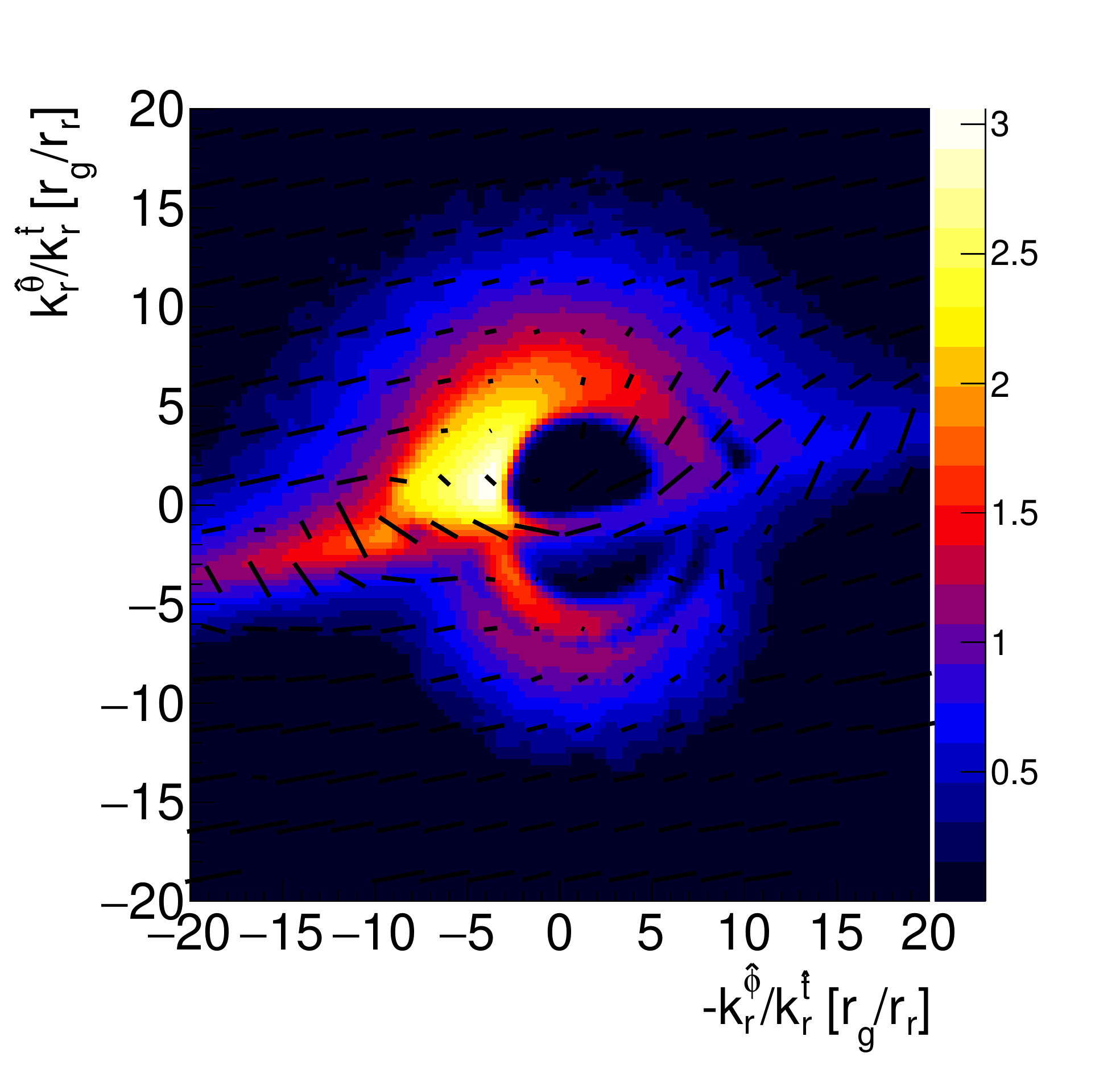}{0.25\linewidth}{}
              \fig{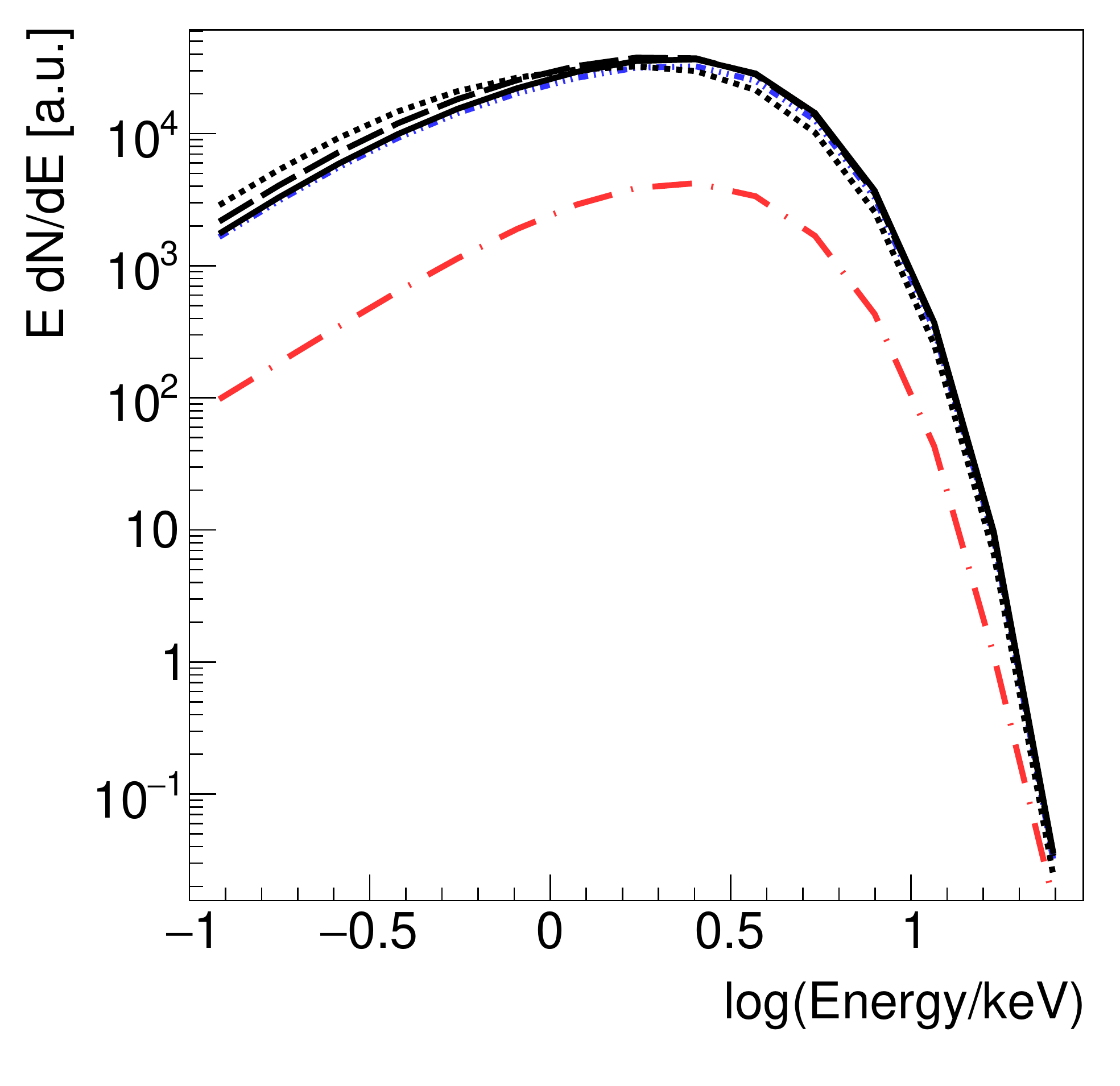}{0.25\linewidth}{}
              \fig{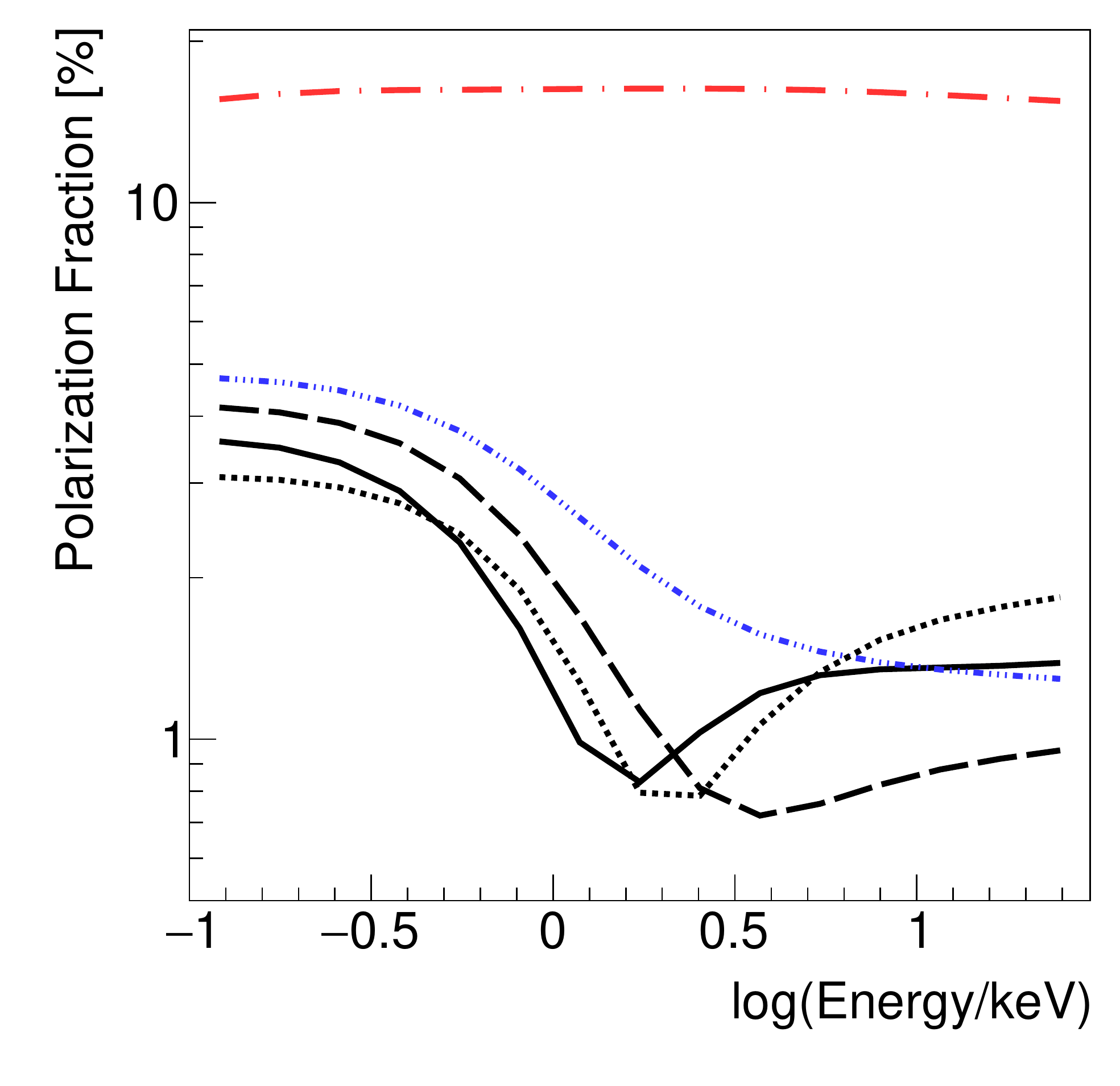}{0.25\linewidth}{}
              \fig{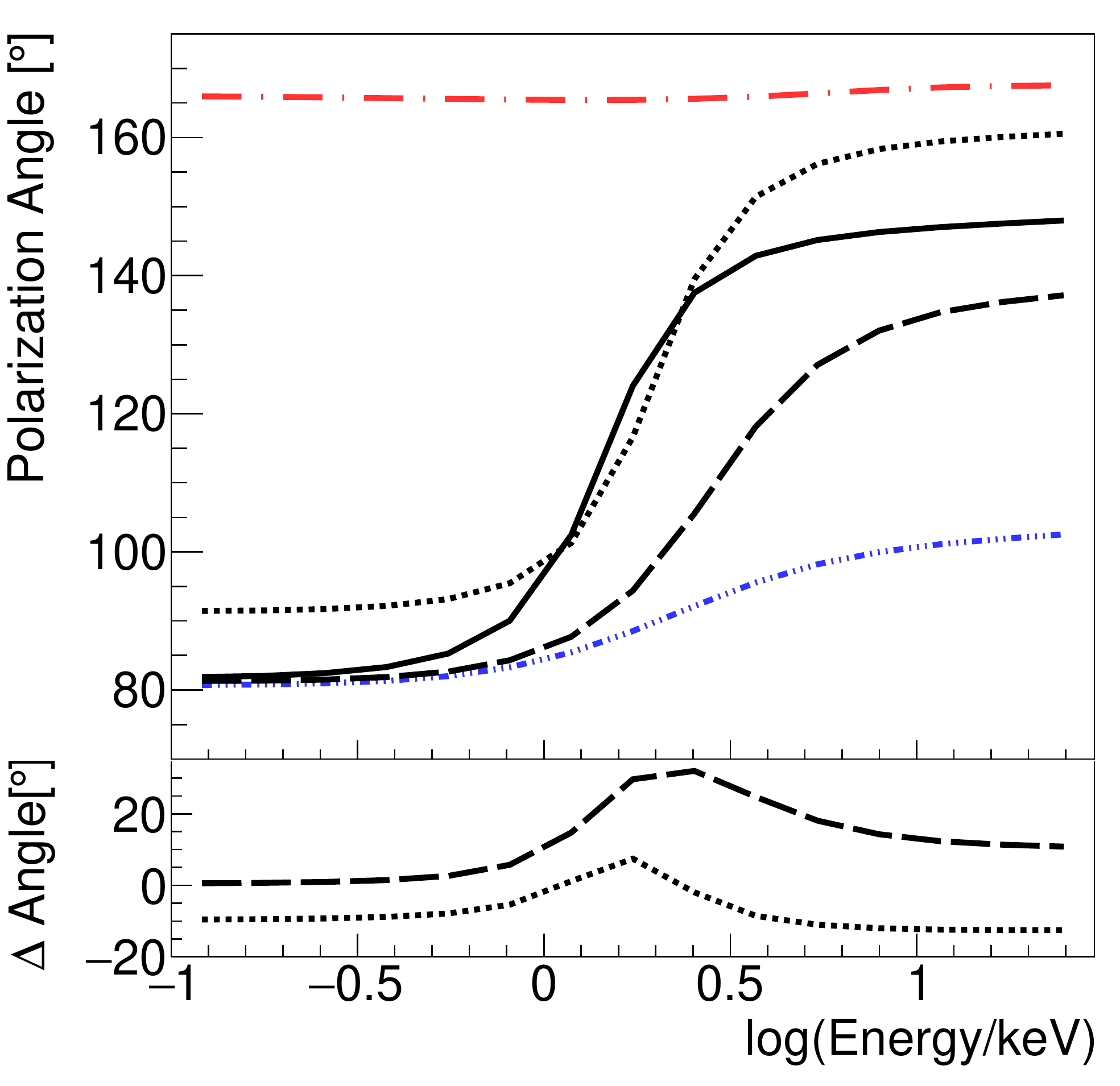}{0.25\linewidth}{}}
    \caption{As in Figure \ref{fig:phi135}, but for \ob{315}.}
    \label{fig:phi315}
\end{figure*}

\ob{315} (Fig.\ \ref{fig:phi315}) sees the outer disk almost edge on; thus it sees fewer photons from the outer disk.
The light curved around the bottom of the BH is visible, which is not true of the completely aligned disk. 
This means the warped disk produces a slightly harder spectrum than the completely aligned disk.
The polarization fraction of the warped disk fits squarely between two flat disk.
Polarization angle even further deviates from either 
flat disk at high energies; it appears to carry the difference between the two flat disks at low energies up past the swing, presumably because many of the high energy photons emitted from the inner disk are scattering off the inclined outer disk instead.
This gives us a clear indication that the polarization angle does just always line up with either that of the inner or outer disks.

\begin{figure*}[t]
    \gridline{\fig{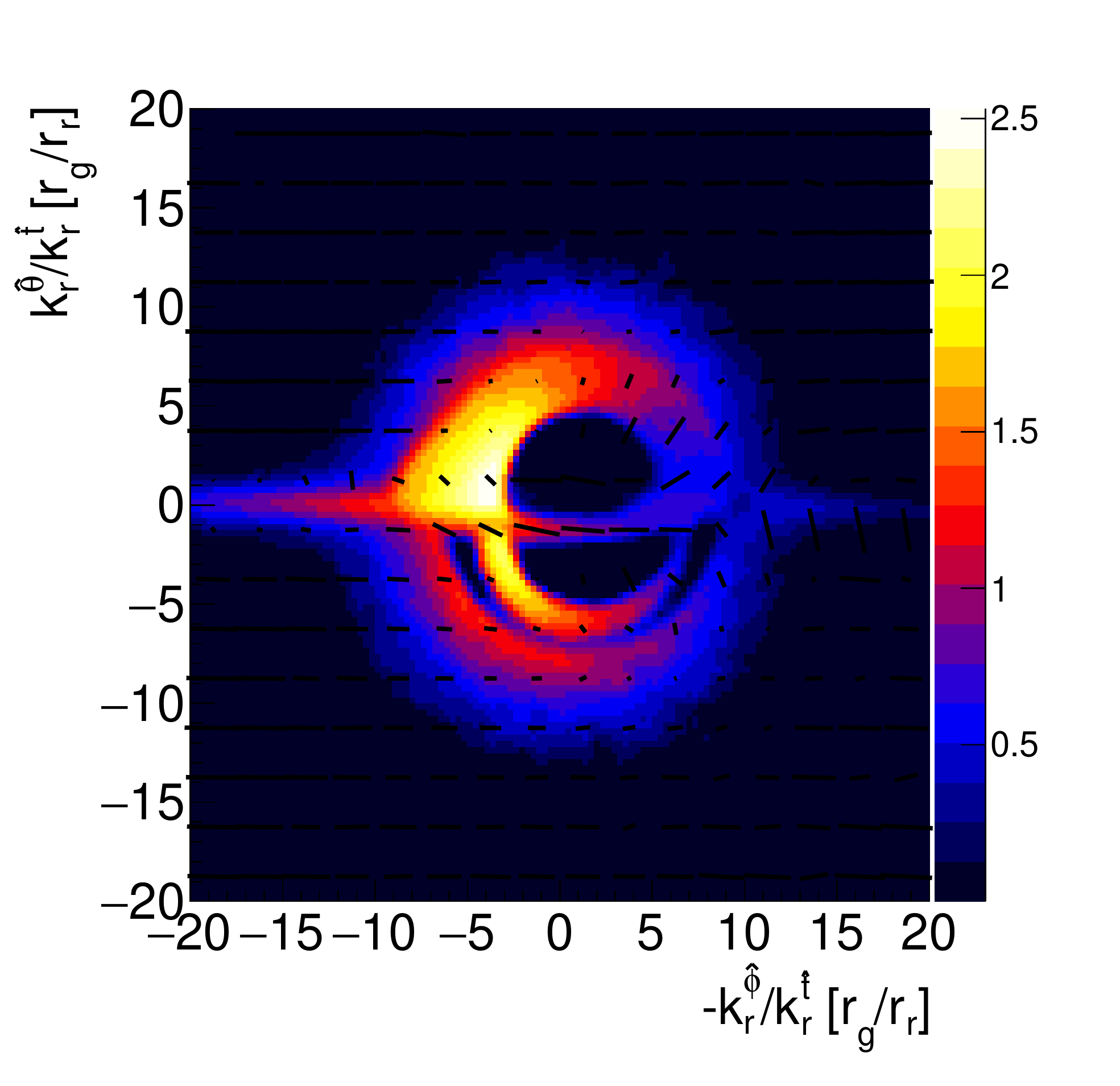}{0.25\linewidth}{}
              \fig{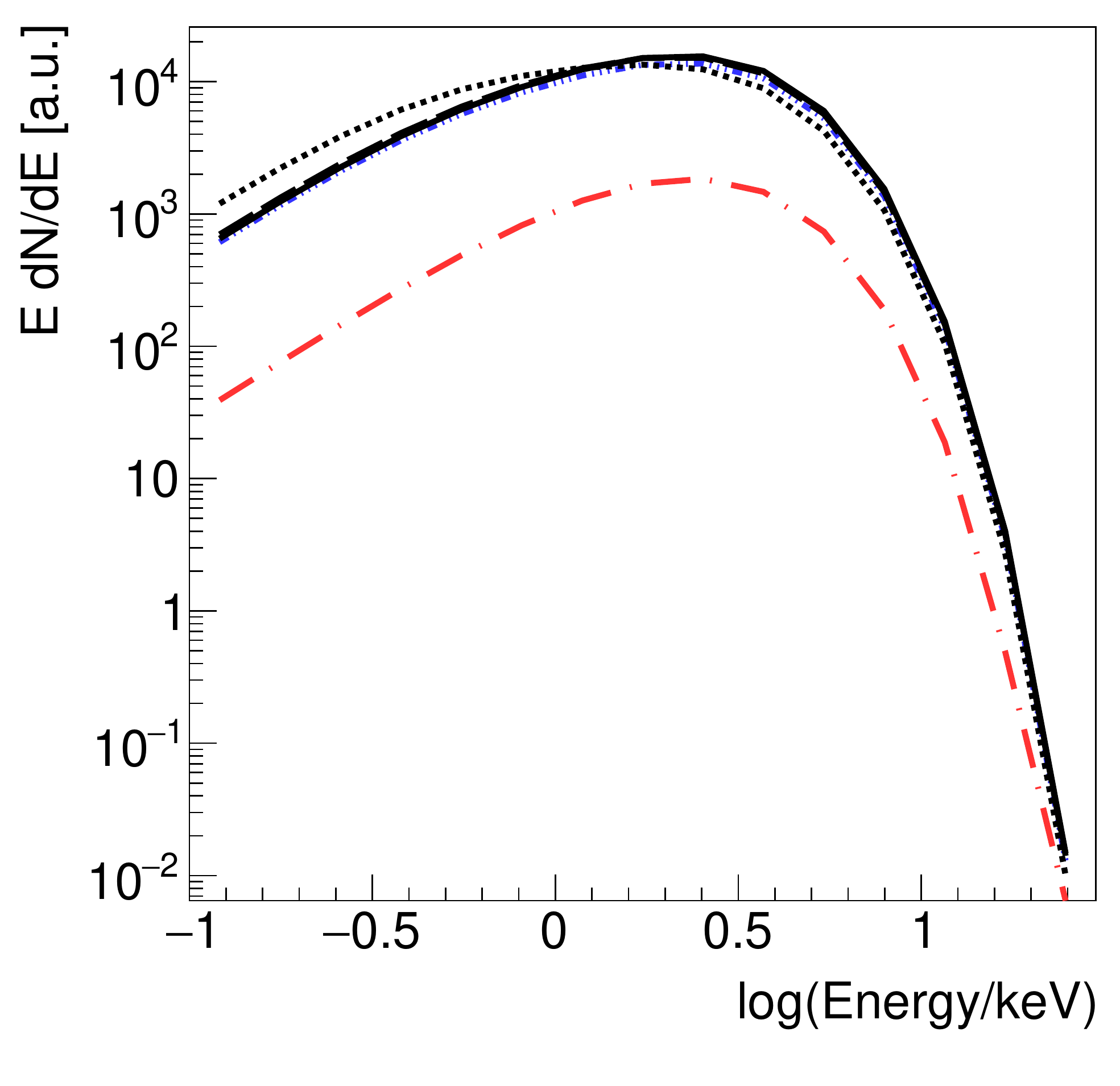}{0.25\linewidth}{}
              \fig{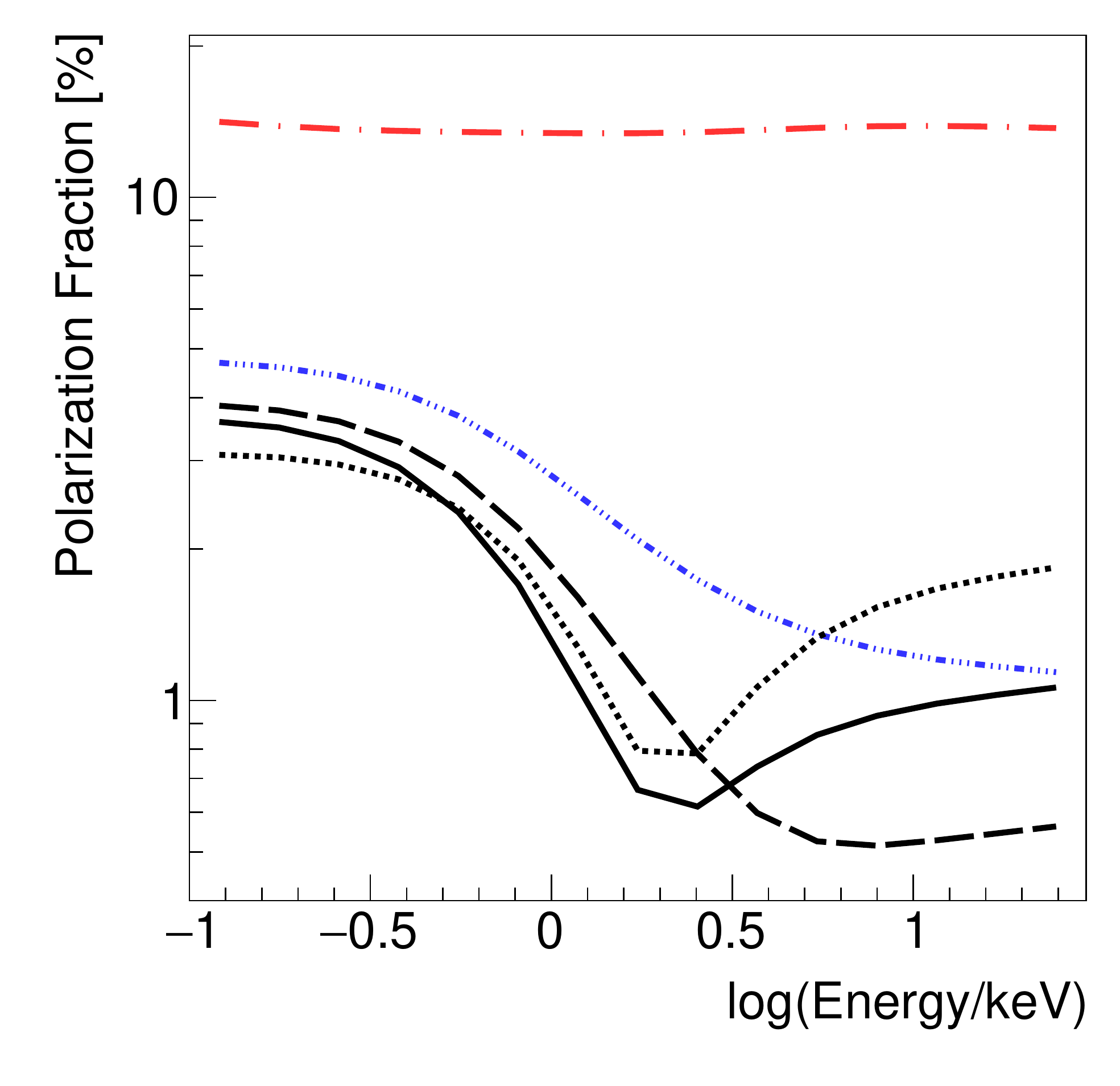}{0.25\linewidth}{}
              \fig{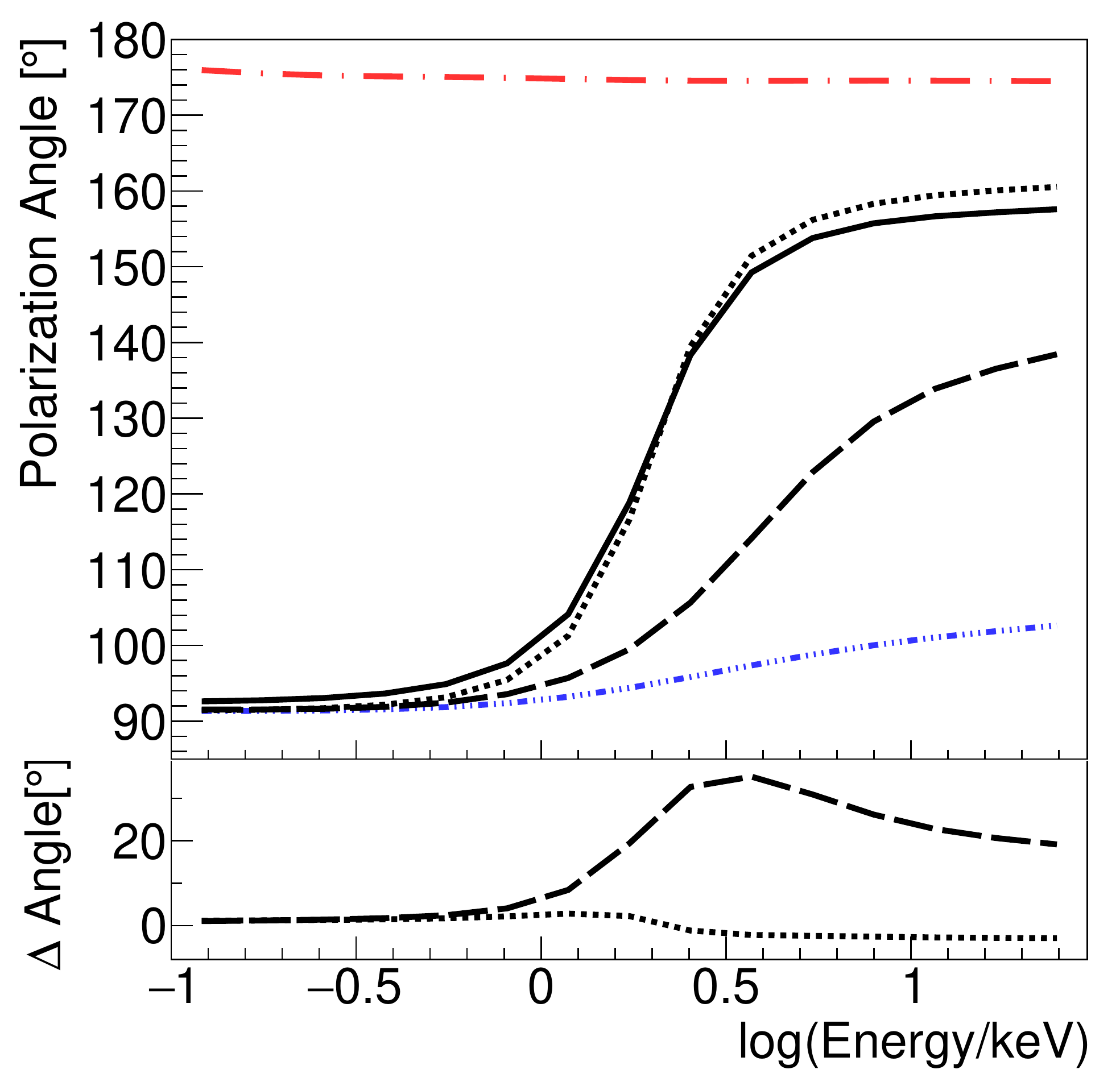}{0.25\linewidth}{}}
    \caption{As in Figure \ref{fig:phi135}, but for \ob{0}.}
    \label{fig:phi0}
\end{figure*}

\ob{0} (Fig.\ \ref{fig:phi0}) sees the outer disk fully edge-on, and thus is somewhat qualitatively similar to \ob{315} (Fig.\ \ref{fig:phi315}).
The polarization angle, though, fully matches the inner disk, just like \ob{180} (Fig.\ \ref{fig:phi180}) who also views no rotation between the inner and outer disks.

\subsection{Comparison to Earlier Results}
Our results can be compared to Figure 5 of \citet{cheng2016} showing the polarization of disks with \rbp-values between \SIrange[range-phrase={ and }]{30}{1000}{\rg}. Their inner disk is inclined at \ang{75}, matching ours, while their inner disk is at \ang{45}; this corresponds to a \eq{$\beta$}{\ang{30}}.
They ignore radiation which is curved by the BH, returns to disk, and is reflected, as well as radiation emitted from the outer disk and reflected by the inner disk.
They find that the disk warp does not affect the polarization angle, but that it is imprinted on a transition in the polarization fraction from matching the inclination of the outer disk at low energies to matching the inner disk at high energies, with the transition changing with \rbp. 

We show instead that in general the polarization does not behave so predictably.
At low energies, where the polarization is dominated by the direct emission from the outer disk, the polarization angle matches the case of the completely misaligned disk.
Polarization fraction at high energies either matches or is lower than that of the completely aligned disk, while it can be higher or lower than the completely misaligned disk depending on the observer.
Polarization angle tends to be offset from the completely aligned disk by the rotation between the two disk axes. 
When there is no visible rotation between the axes of the outer and inner parts of the warped disk (i.e. all of the misalignment is in the inclination difference), the polarization angle matches the completely aligned disk.
For every other observer, the polarization angle roughly matches the completely misaligned disks until the observer sees the warped outer disk close to edge on; in these cases, the scattered X-rays at high energies carry the angle of rotation between the inner and outer disks.

Thus, with some measure of the jet direction, the polarization angle can give a lower limit on the misalignment between the inner and outer disks.
\citet{cheng2016} note that the energy at which the polarization fraction transitions from that of the outer disk inclination to the inner disk inclination is inversely related to \rbp.
Since the closest of our observers to theirs is \ob{180} (Fig.\ \ref{fig:phi180}), where we see the polarization fraction is lower across the board, we assume that the transition energy corresponding to \eq{\rbp}{\SI{8}{\rg}} is greater than the energy range of our analysis.
We also note that this is not true for other observers, where the misalignment is not entirely measured by the difference in inclinations between the inner and outer warped disks.

\begin{figure}
    \centering
    \includegraphics[width=0.8\linewidth]{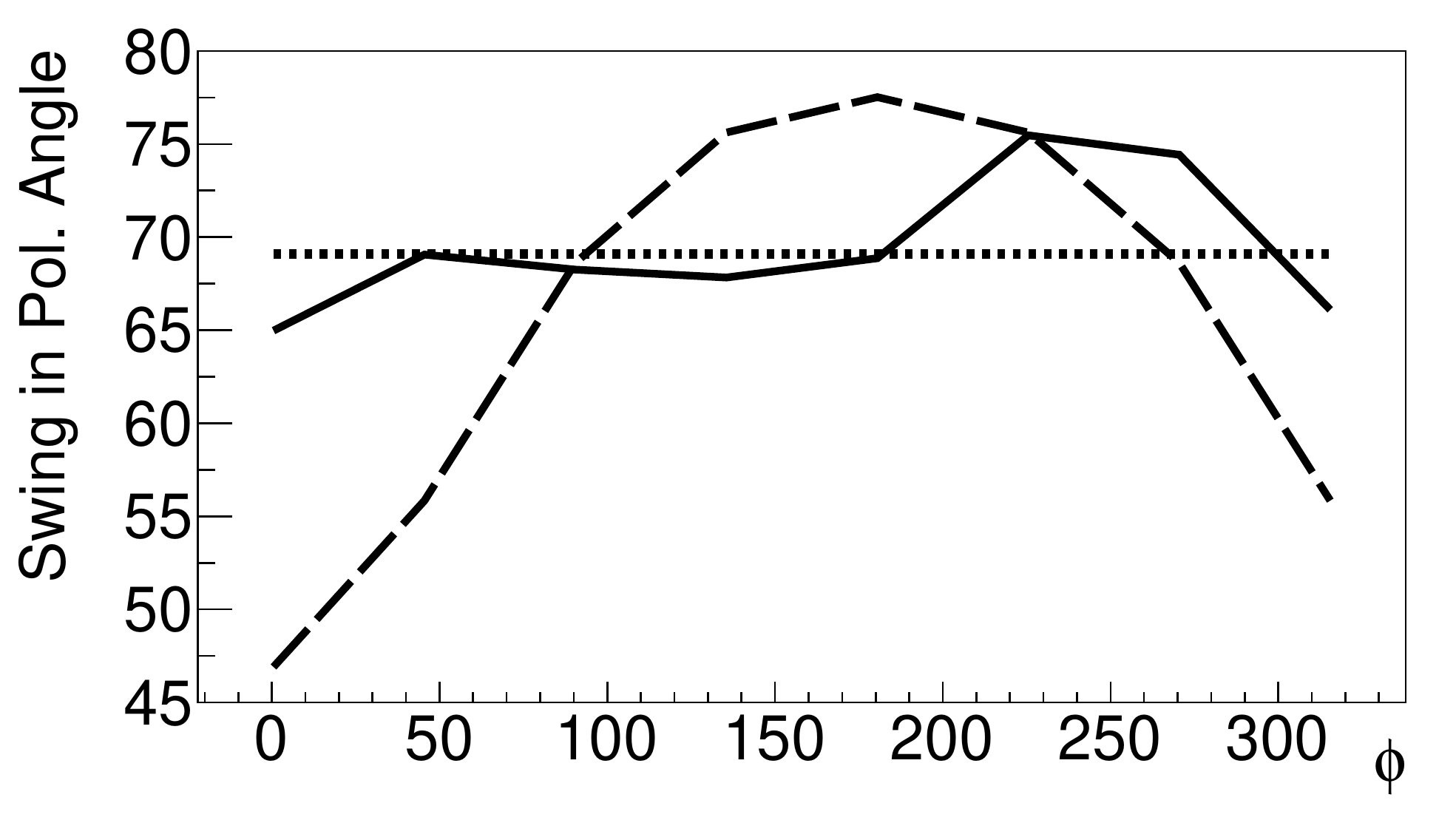}
    \caption{The swing in polarization angle over this inclination. Shown are for the warped disk (solid black), the completely aligned disk (dotted), and completely misaligned disk (dashed).}
    \label{fig:swing}
\end{figure}
To see how these observers might be distinguished, we plotted in Figure \ref{fig:swing} the swing in polarization angle for each warped disk, completely misaligned disk, and the completely aligned disk. This shows us that the warped disk at this inclination has a swing that could always be mistaken for a flat disk. 

\section{Discussion \& Conclusions} \label{sec:discuss}

In this paper, we examine the spectropolarimetric signature of a warped accretion disk, similar to what might result from the Bardeen-Petterson effect. 
We find that small values of \rbp\, complicate the simple
picture that the polarization fraction and direction at high (low) energies are mostly determined by the inner (outer)
accretion disk properties.
Since most returning radiation scatters within \SI{\sim 10}{\rg}, a small value for \rbp means that some of this radiation reflects off the outer disk.
Based on Figure \ref{fig:images90}, though, it does appear that simply the presence of a warped disk means that more returning radiation scatters at larger radii.
Our results show that the measured polarization is highly dependent on the azimuthal viewing angle of the misaligned system, especially in regards to the polarization angle.

Since a warped disk does not affect the energy at which the polarization angle swings, polarization can still be used to measure BH spin (see Fig. 7 in \citet{sk2009}). 
Similarly, the continuum-fitting method \citep{mcclintock2014} is still valid as the spectrum is not significantly modified by the presence of a warped disk.
The complication to using the continuum-fitting method is instead in assuming the inclination of the inner accretion disk based on eclipse observations (though jet inclinations are still valid).
For a given source, then, polarimetric observations could tell us whether the binary plane inclination is appropriate assumption.

For systems like the aforementioned GRO J1655-40 where we can get one or both of the jet and binary inclinations, polarization can be a powerful tool to examine the overall alignment.
The jet inclination of GRO J1655-40 was measured as \ang{85} \citep{hjellming1995} and the binary inclination as \ang{70} \citep{greene2001}.
By these two measurements, the misalignment is \ang{\ge15}; we could verify this by seeing if the polarization angle aligns with the jet.
We may also expect to measure a lower misalignment with polarization, as \citet{liska2019a} showed that global alignment can make the disk warped smaller than the overall system misalignment.

Cygnus X-1, one of the most famous and well-studied black hole binaries, may also be misaligned, and thus is one of the best candidates to study disk warping with X-ray polarization.
The inner disk inclination disagrees with the binary plane measured by \citet{orosz2011} by \SIrange{\sim10}{15}{\degree} \citep{tomsick2014, walton2016}.
Measuring the disk warp would also help fit reverberation measurements \citep[e.g.]{mastroserio2019}, allowing for more accurate independent measures of the black hole mass.

Another possible target is 4U 1957+11, a low-mass X-ray binary that is a black hole candidate.
It is consistently in the spectrally soft state, and is likely highly spinning and highly inclined \citep{nowak2012}.
Most of the time it is well fit by a purely thermal spectrum, so would be a good test case for the results in this paper.

There is some concern that without a prior measurement of the system orientation, measurements of the polarization angle will be relatively useless.
We are fortunate, then, that misaligned accretion is thought to be a driver of corona and jet formation in BH binaries \citep{king2018}, and so an observed jet is an indication that we should target the source for polarimetric measurement.
Another benifit to focusing on sources with visible jets is their association with a strong power-law component; we expect that the time lag in the reflected power-law emission from the corona will track \rbp and the misalignment; this is the subject of a future publication (Abarr et al. 2019, in preparation).

The Bardeen-Petterson effect has many interesting implications on the dynamics of accreting BHs, and polarimetry is in a unique position to examine this.
X-ray polarimetry could confirm that the precession of the inner disk before its angular momentum is aligned with the BH is the driver of quasi-periodic oscillations \citep{ingram2015}.
The BP effect could help explain state transitions in BH X-ray binaries according to \citet{nixon2014}.
In their model, a warp or break in the disk is present in both the soft-intermediate state (SIMS) and high/soft state (HSS), so we would expect to see some polarimetric signature of these, possibly of both the thermal and power-law emission (which becomes prominent in the SIMS).

With the working basis for an inclined outer disk we have described in this paper (Eq. \ref{eq:bpbasis}), many of these ideas can be explored in future projects. This will be essential for interpretation of the upcoming results from IXPE, XL-Calibur, and other future X-ray polarimetric observatories.

\acknowledgments

Q.A. would like to thank Augusto Madeiros da Rosa for invaluable guidance. The authors thank NASA for the support through the grants NNX16AC42G and 80NSSC18K0264, as well as the Washington University McDonnell Center for the Space Sciences.


\begin{thebibliography}{}

\bibitem[Bardeen \& Petterson(1975)]{BP1975} Bardeen, J.~M., \& Petterson, J.~A.\ 1975, \apjl, 195, L65
\bibitem[Bardeen et al.(1972)]{bardeen1972} Bardeen, J.~M., Press, W.~H., \& Teukolsky, S.~A.\ 1972, \apj, 178, 347
\bibitem[Beheshtipour et al.(2017)]{beheshtipour2017} Beheshtipour, B., Krawczynski, H., \& Malzac, J.\ 2017, \apj, 850, 14
\bibitem[Cash \& Karp(1990)]{ck1990} Cash, J.R., \& Karp, A.H.\ 1990, ACM Transactions on Mathematical Software 16, 201
\bibitem[Chandrasekhar(1960)]{chandrasekhar} Chandrasekhar, S.\ 1960, New York: Dover
\bibitem[Cheng et al.(2016)]{cheng2016} Cheng, Y., Liu, D., Nampalliwar, S., et al.\ 2016, Classical and Quantum Gravity, 33, 125015
\bibitem[Fragile et al.(2001)]{fragile2001} Fragile, P.~C., Mathews, G.~J., \& Wilson, J.~R.\ 2001, \apj, 553, 955
\bibitem[Greene et al.(2001)]{greene2001} Greene, J., Bailyn, C.~D., \& Orosz, J.~A.\ 2001, \apj, 554, 1290
\bibitem[Hjellming, \& Rupen(1995)]{hjellming1995} Hjellming, R.~M., \& Rupen, M.~P.\ 1995, \nat, 375, 464
\bibitem[Hoormann et al.(2016)]{Hoor:16} Hoormann, J.~K., Beheshtipour, B., \& Krawczynski, H.\ 2016, \prd, 93, 044020
\bibitem[Ingram et al.(2015)]{ingram2015} Ingram, A., Maccarone, T.~J., Poutanen, J., et al.\ 2015, \apj, 807, 53
\bibitem[King et al.(2005)]{king2005} King, A.~R., Lubow, S.~H., Ogilvie, G.~I., et al.\ 2005, \mnras, 363, 49
\bibitem[King, \& Nixon(2018)]{king2018} King, A., \& Nixon, C.\ 2018, The Astrophysical Journal, 857, L7
\bibitem[Krawczynski(2012)]{krawcz2012} Krawczynski, H.\ 2012, \apj, 754, 133
\bibitem[Krawczynski et al.(2019)]{krawcz2019} Krawczynski, H., Chartas, G., \& Kislat, F.\ 2019, \apj, 870, 125
\bibitem[Lense, \& Thirring(1918)]{lt1918} Lense, J., \& Thirring, H.\ 1918, Physikalische Zeitschrift, 19, 156
\bibitem[Liska et al.(2018)]{liska2018} Liska, M., Hesp, C., Tchekhovskoy, A., et al.\ 2018, \mnras, 474, L81
\bibitem[Liska et al.(2019)]{liska2019a} Liska, M., Tchekhovskoy, A., Ingram, A., et al.\ 2019, \mnras, 487, 550
\bibitem[Mastroserio et al.(2019)]{mastroserio2019} Mastroserio, G., Ingram, A., \& van der Klis, M.\ 2019, \mnras, 488, 348
\bibitem[McClintock et al.(2014)]{mcclintock2014} McClintock, J.~E., Narayan, R., \& Steiner, J.~F.\ 2014, \ssr, 183, 295
\bibitem[McKinney et al.(2013)]{mckinney2013} McKinney, J.~C., Tchekhovskoy, A., \& Blandford, R.~D.\ 2013, Science, 339, 49
\bibitem[Morales Teixeira et al.(2018)]{morales2018} Morales Teixeira, D., Avara, M.~J., \& McKinney, J.~C.\ 2018, \mnras, 480, 3547
\bibitem[Nixon, \& King(2012)]{nixon2012a} Nixon, C.~J., \& King, A.~R.\ 2012, \mnras, 421, 1201
\bibitem[Nixon et al.(2012)]{nixon2012b} Nixon, C., King, A., Price, D., et al.\ 2012, \apjl, 757, L24
\bibitem[Nixon, \& Salvesen(2014)]{nixon2014} Nixon, C., \& Salvesen, G.\ 2014, \mnras, 437, 3994
\bibitem[Nowak et al.(2012)]{nowak2012} Nowak, M.~A., Wilms, J., Pottschmidt, K., et al.\ 2012, \apj, 744, 107
\bibitem[Ogilvie(1999)]{ogilvie1999} Ogilvie, G.~I.\ 1999, \mnras, 304, 557
\bibitem[Orosz et al.(2011)]{orosz2011} Orosz, J.~A., McClintock, J.~E., Aufdenberg, J.~P., et al.\ 2011, \apj, 742, 84
\bibitem[Page \& Thorne(1974)]{Page:74} Page, D.~N., \& Thorne, K.~S.\ 1974, \apj, 191, 499 
\bibitem[Remillard, \& McClintock(2006)]{rm2006} Remillard, R.~A., \& McClintock, J.~E.\ 2006, \araa, 44, 49
\bibitem[Risaliti et al.(2013)]{risaliti2013} Risaliti, G., Harrison, F.~A., Madsen, K.~K., et al.\ 2013, \nat, 494, 449
\bibitem[Schnittman, \& Krolik(2009)]{sk2009} Schnittman, J.~D., \& Krolik, J.~H.\ 2009, \apj, 701, 1175
\bibitem[Schnittman, \& Krolik(2013)]{schnittman2013} Schnittman, J.~D., \& Krolik, J.~H.\ 2013, \apj, 777, 11
\bibitem[Sorathia et al.(2013)]{sorathia2013} Sorathia, K.~A., Krolik, J.~H., \& Hawley, J.~F.\ 2013, \apj, 777, 21
\bibitem[Tomsick et al.(2014)]{tomsick2014} Tomsick, J.~A., Nowak, M.~A., Parker, M., et al.\ 2014, \apj, 780, 78
\bibitem[Walton et al.(2016)]{walton2016} Walton, D.~J., Tomsick, J.~A., Madsen, K.~K., et al.\ 2016, \apj, 826, 87
\bibitem[Weisskopf et al.(2016)]{weisskopf2016} Weisskopf, M.~C., Ramsey, B., O'Dell, S., et al.\ 2016, \procspie, 990517
\bibitem[Wilkins(1972)]{wilkins72} Wilkins, D.~C.\ 1972, \prd, 5, 814

\end{thebibliography}
\end{document}